\documentclass[aps,twocolumn,showpacs]{revtex4}
\usepackage{amsmath}
\usepackage{epsfig}
\usepackage{multirow}
\usepackage{appendix}
\usepackage{amssymb}

\begin{document}
	
\title{Prediction of $P_{cc}$ states in quark model}
\author{Ye Yan$^1$}\email{221001005@njnu.edu.cn}
\author{Hongxia Huang$^1$}\email{hxhuang@njnu.edu.cn(Corresponding author)}
\author{Xinmei Zhu$^2$}\email[E-mail: ]{xmzhu@yzu.edu.cn(Corresponding author)}
\author{Jialun Ping$^1$}\email{jlping@njnu.edu.cn}
\affiliation{$^1$School of Physics and Technology, Nanjing Normal University, Nanjing 210097, People's Republic of China}
\affiliation{$^2$Department of Physics, Yangzhou University, Yangzhou 225009, People's Republic of China}

\begin{abstract}
Inspired by the observation of hidden-charm pentaquark $P_c$ and $P_{cs}$ states by the LHCb Collaboration, we explore the $qqc\bar{c}c$ ($q~=~u$ or $d$) pentaquark systems in the quark delocalization color screening model.
The interaction between baryons and mesons and the influence of channel coupling are studied in this work.
Three compact $qqc\bar{c}c$ pentaquark states are obtained, whose masses are 5259 MeV with $I(J^P)$ = $0(1/2^-)$, 5396 MeV with $I(J^P)$ = $1(1/2^-)$, and 5465 MeV with $I(J^P)$ = $1(3/2^-)$.
Two molecular states are obtained, which are $I(J^P)$ = $0(1/2^-)$ $\Lambda_c J/\psi$ with 5367 MeV and $I(J^P)$ = $0(5/2^-)$ $\Xi_{cc}^* \bar{D}^*$ with 5690 MeV.
These predicted states may provide important information for future experimental search.
\end{abstract}

	
\maketitle

\setcounter{totalnumber}{5}
	
\section{\label{sec:introduction}Introduction}

In the recent years, regarding the exotic hadron studies, one of the most noteworthy experimental progresses is the observation of the hidden-charm pentaquark states, which brings great interest in pentaquark investigations.
In 2015, the LHCb Collaboration reported two states $P_{c}(4380)$ and $P_{c}(4450)$ in the $J/\psi p$ invariant mass spectrum of $\Lambda^{0}_{b} \rightarrow J/\psi K^{-}p$~\cite{LHCbPc2015}.
Subsequently, the LHCb Collaboration updated their results in 2019, a new states $P_{c}(4312)$ was proposed, and the $P_{c}(4450)$ was split to $P_{c}(4440)$ and $P_{c}(4457)$~\cite{LHCbPc2019}.

Before the LHCb collaboration's observations, several theoretical works have predicted the existence of the hidden-charm pentaquark states~\cite{Wu:2010jy,Wang:2011rga,Yang:2011wz,Xiao:2013yca}.
After the experimental report, a wide range of theoretical researches on these $P_{c}$ states were initiated~\cite{Chen:2015moa,Wang:2015epa,Wang:2015ava,Azizi:2016dhy,Azizi:2018bdv,Chen:2019bip,Zhang:2019xtu,Xu:2019zme,Wang:2019got,Azizi:2020ogm,Xu:2020flp,Ozdem:2018qeh,Lu:2015fva,Lu:2016nnt,Xiao:2019mvs,Wu:2019rog,Wang:2019dsi,Zhang:2020erj,Xu:2020gjl,Li:2023zag,Roca:2015dva,Shimizu:2016rrd,Meng:2019ilv,He:2015cea,He:2019ify,Oset:2016nvf,Wang:2019ato,Peng:2020xrf,Wang:2015qlf,Ali:2016dkf,Shimizu:2019ptd,Du:2021fmf,Guo:2015umn,Liu:2015fea,Guo:2016bkl,Santopinto:2016pkp,Ortega:2016syt,Park:2017jbn,Zhu:2019iwm,Phumphan:2021tta,Yang:2015bmv,Dong:2020nwk,Chen:2015loa,Liu:2019zvb,Chen:2019asm,Yalikun:2021bfm,Deng:2016rus,HuangPc1,HuangPc2,Zhu:2023hyh,Scoccola:2015nia,Wu:2017weo,Weng:2019ynv,Zhu:2015bba,Wang:2016dzu,Hiyama:2018ukv,Burns:2019iih,Yang:2022ezl,Meissner:2015mza,Mutuk:2019snd,Yamaguchi:2019seo,Kubarovsky:2015aaa,Wang:2015jsa,HillerBlin:2016odx,Paryev:2018fyv,Guo:2019fdo,Shen:2016tzq,Burns:2015dwa}.
These states have been investigated using the QCD (Quantum Chromodynamics) sum rules~\cite{Chen:2015moa,Wang:2015epa,Wang:2015ava,Azizi:2016dhy,Azizi:2018bdv,Chen:2019bip,Zhang:2019xtu,Xu:2019zme,Wang:2019got,Azizi:2020ogm,Xu:2020flp}, light-cone QCD sum rules~\cite{Ozdem:2018qeh},
effective Lagrangian approach~\cite{Lu:2015fva,Lu:2016nnt,Xiao:2019mvs,Wu:2019rog,Wang:2019dsi,Zhang:2020erj,Xu:2020gjl,Li:2023zag},
chiral Lagrangian approach~\cite{Roca:2015dva,Shimizu:2016rrd},
chiral perturbation theory~\cite{Meng:2019ilv},
quasipotential Bethe–Saltpeter approach~\cite{He:2015cea,He:2019ify},
chiral unitary approach~\cite{Oset:2016nvf},
effective field theory~\cite{Wang:2019ato,Peng:2020xrf},
heavy quark spin symmetry~\cite{Wang:2015qlf,Ali:2016dkf,Shimizu:2019ptd,Du:2021fmf},
triangle singularity~\cite{Guo:2015umn,Liu:2015fea,Guo:2016bkl}, G\"ursey-Radicati inspired mass formula~\cite{Santopinto:2016pkp},
constituent quark model~\cite{Ortega:2016syt,Park:2017jbn,Zhu:2019iwm,Phumphan:2021tta,Yang:2015bmv,Dong:2020nwk,HuangPc1,HuangPc2,Zhu:2023hyh},
one-boson exchange model~\cite{Chen:2015loa,Liu:2019zvb,Chen:2019asm,Yalikun:2021bfm},
color flux-tube model~\cite{Deng:2016rus},
topological soliton model~\cite{Scoccola:2015nia},
color-magnetic interaction model~\cite{Wu:2017weo,Weng:2019ynv},
molecular models~\cite{Zhu:2015bba,Wang:2016dzu,Hiyama:2018ukv,Burns:2019iih,Yang:2022ezl},
and various other models~\cite{Meissner:2015mza,Mutuk:2019snd,Yamaguchi:2019seo}.
In addition, the production mechanism of these states is also investigated in Refs.~\cite{Kubarovsky:2015aaa,Wang:2015jsa,HillerBlin:2016odx,Paryev:2018fyv}.
Given the near-threshold nature of the $P_c$ states, most of the theoretical works interpret them as molecular states.
For instance, in Ref.~\cite{Chen:2019asm}, the authors performed a direct calculation using the one-boson exchange model and explicitly demonstrated that the $P_{c}(4312)$, $P_{c}(4440)$ and $P_{c}(4457)$ correspond to the loosely bound $\Sigma_c \bar{D}$ with ($I$ = 1/2, $J^P = 1/2^-$), $\Sigma_c \bar{D}^*$ with ($I$ = 1/2, $J^P = 1/2^-$), and $\Sigma_c \bar{D}^*$ with ($I$ = 1/2, $J^P= 3/2 ^-$ ), respectively.
More detailed reviews on the $P_{c}$ states can also be found in Refs.~\cite{ChenHX0,LiuYR0,YangG0}.

In 2020, the LHCb Collaboration reported a 3$\sigma$ hidden-charm strange pentaquark structure $P_{cs}(4459)$ in the $\Xi^-_b \rightarrow J/\psi \Lambda K^-$ decay~\cite{LHCb:2020jpq}.
In 2022, the LHCb Collaboration reported their results about the $B^- \rightarrow J/\psi \Lambda \bar{p}$ decay, which indicates the existence of a new hidden-charm strange pentaquark state $P_{cs}(4338)$~\cite{LHCb:2022ogu}.
Given that the mass of the $P_{cs}(4459)$ and $P_{cs}(4338)$ is close to the threshold of $\Xi_c \bar{D}^*$ and $\Xi_c \bar{D}$, respectively, it has led to various theoretical studies on the $P_{cs}$ states\cite{Chen:2020uif,Wang:2020eep,Azizi:2021utt,Wang:2021itn,Wang:2022neq,Azizi:2023iym,Ozdem:2021ugy,Ozdem:2022kei,Ozdem:2023htj,Peng:2020hql,Liu:2020hcv,Yan:2021nio,Yan:2022wuz,Cheng:2021gca,Yang:2021pio,Wu:2021caw,Clymton:2021thh,Zhu:2021lhd,Zhu:2022wpi,Burns:2022uha,Feijoo:2022rxf,Xiao:2021rgp,Chen:2020kco,Chen:2021tip,Wang:2022mxy,Chen:2022onm,Li:2023aui,Hu:2021nvs,Deng:2022vkv,Wang:2022tib,Ortega:2022uyu,Gao:2021hmv,Meng:2022wgl,Giachino:2022pws,Paryev:2022zdx,Chen:2022wkh,Paryev:2023icm,Nakamura:2022gtu,Liu:2020ajv}.
These states have been investigated in the framework of the QCD sum rules~\cite{Chen:2020uif,Wang:2020eep,Azizi:2021utt,Wang:2021itn,Wang:2022neq,Azizi:2023iym},
light-cone QCD sum rules~\cite{Ozdem:2021ugy,Ozdem:2022kei,Ozdem:2023htj},
effective field theory~\cite{Peng:2020hql,Liu:2020hcv,Yan:2021nio,Yan:2022wuz},
effective Lagrangian approach~\cite{Cheng:2021gca,Yang:2021pio,Wu:2021caw,Clymton:2021thh},
quasipotential Bethe-Salpeter equation approach~\cite{Zhu:2021lhd,Zhu:2022wpi},
triangle singularity~\cite{Burns:2022uha},
heavy quark spin symmetry~\cite{Feijoo:2022rxf},
coupled channel unitary approach~\cite{Xiao:2021rgp},
constituent quark model~\cite{Wang:2022tib,Ortega:2022uyu,Gao:2021hmv,Hu:2021nvs},
one-boson exchange model~\cite{Chen:2020kco,Chen:2021tip,Wang:2022mxy,Chen:2022onm}, the color-magnetic interaction model~\cite{Li:2023aui},
color flux-tube model~\cite{Deng:2022vkv},
zero-range model and the Flatté model~\cite{Meng:2022wgl}, and other models~\cite{Giachino:2022pws,Paryev:2022zdx,Chen:2022wkh,Paryev:2023icm,Nakamura:2022gtu}.
One of the focuses of these theoretical works is to determine the structure of the $P_{cs}$ states.
For the $P_{cs}(4459)$, conclusion of molecular configuration is supported in Refs.~\cite{Chen:2020uif,Wang:2021itn,Peng:2020hql,Yan:2022wuz,Cheng:2021gca,Yang:2021pio,Wu:2021caw,Xiao:2021rgp,Zhu:2021lhd,Feijoo:2022rxf,Chen:2020kco,Chen:2021tip,Wang:2022mxy,Hu:2021nvs,Ortega:2022uyu}, while there are also theoretical works that interpret the $P_{cs}(4459)$ as a compact pentaquark~\cite{Wang:2020eep,Azizi:2021utt,Li:2023aui,Deng:2022vkv}.
For the $P_{cs}(4338)$, molecular configuration is preferred in Refs.~\cite{Wang:2022neq,Azizi:2023iym,Yan:2022wuz,Zhu:2022wpi,Feijoo:2022rxf,Wang:2022mxy,Giachino:2022pws,Ortega:2022uyu}.
However, it can be interpreted as $udsc\bar{c}$ compact pentaquark state according to Ref.~\cite{Li:2023aui}.
Determining the structure of different states in the same theoretical framework is a meaningful subject.
According to the QCD sum rules results of Refs.~\cite{Azizi:2021utt,Azizi:2023iym}, the compact pentaquark nature of diquark-diquark-antiquark form with $J^P = 1/2^-$ is favored for the $P_{cs}(4459)$ state, and the $\Xi_c \bar{D}$ molecular nature with $J^P = 1/2^-$ is favored for the $P_{cs}(4338)$ state.

Given the existence of the $P_{c}$ $(qqqc\bar{c})$ and $P_{cs}$ $(qqsc\bar{c})$ states, one may wonder if there are other types of pentaquark states that contain a pair of $c\bar{c}$.
If the strange quark in the $P_{cs}$ state is replaced by the charm quark, it is interesting to explore if there exists the $P_{cc}$ state $(qqcc\bar{c})$.
Additionally, according to Ref.~\cite{Gershon:2022xnn}, $P_{cc}$ state can also be named $P_{\psi c}^{\Lambda}$ and $P_{\psi c}^{\Sigma}$ based on the isospin.
Several theoretical studies have been carried out concerning the $(qqqc\bar{c})$ system~\cite{Hofmann:2005sw,Chen:2017jjn,Wang:2018ihk,Wang:2019aoc,An:2019idk}.
In Ref.~\cite{Chen:2017jjn,Wang:2019aoc}, the $\Xi_{cc} \bar{D}$, $\Xi_{cc} \bar{D}^*$, $\Xi_{cc} \bar{D}_1$, and $\Xi_{cc} \bar{D}_2^*$ interactions are studied from molecular picture by using the one-boson exchange model, and several possible states were found, which are $\Xi_{cc} \bar{D}^*$ with $I(J^P) = 0(1/2^-)$, $\Xi_{cc} \bar{D}_1$ with $I(J^P) = 0(1/2^+, 3/2^+)$, and $\Xi_{cc} \bar{D}_2^*$ with $I(J^P) = 0(3/2^+, 5/2^+)$.
In Ref.~\cite{Wang:2018ihk}, the scalar-diquark-scalar-diquark-antiquark type current is constructed to interpolate the $P_{cc\bar{c}ud}$ pentaquark states with $J^P=1/2^\pm$.
In Ref.~\cite{An:2019idk}, in the framework of the color-magnetic interaction model, the $qqcc\bar{c}$ $(q=u, d)$ with $I(J^P) = 0(5/2^-)$ is determined to be a possible stable state.

The quark delocalization color screening model (QDCSM) was developed with the aim of explaining the similarities between nuclear and molecular forces~\cite{Wu:1996fm}.
This model gives a good description of the $NN$ and $YN$ interactions and the properties of deuteron~\cite{Ping:2000dx,Ping:1998si,Wu:1998wu,Pang:2001xx}.
It is also employed to calculate the baryon-baryon and baryon-meson scattering phase shifts, and the exotic hadronic states are also studied in this model~\cite{Huang:2023jec}.
Studies show that color screening is an effective description of the hidden-color channel coupling~\cite{ChenLZ,Huang:2011kf}.
When the LHCb collaboration reported the $P_{c}$ states for the first time, the QDCSM is employed to study this system and seven states are obtained~\cite{HuangPc1,HuangPc2}.
Three of them can be used to explain the updated results of $P_{c}$ states reported by the LHCb collaboration in 2019.
Therefore, it is feasible and meaningful to extend this model to investigate the $qqcc\bar{c}$ pentaquark system.

In this work, we systematically investigate the $qqc\bar{c}c$ pentaquark systems in order to find the possible states.
First, the effective potential is studied to understand the interaction between baryon and meson.
The five-body system is calculated by means of the resonating group method to search for bound states.
The influence of channel coupling is discussed based on the current results.
In addition, we calculate the scattering phase shift to examine the possible resonance states.

This paper is organized as follows.
After introduction, the details of QDCSM are presented in Sec.~II.
The effective potential, the bound-state calculation, and the scattering process are presented in Sec.~III, along with the discussion and analysis of the results.
Finally, the paper ends with a summary in Sec.~IV.

\section{QUARK DELOCALIZATION COLOR SCREENING MODEL (QDCSM)}
Herein, the QDCSM is employed to investigate the properties of $ssc\bar{q}q$ systems.
The QDCSM is an extension of the native quark cluster model~\cite{DeRujula:1975qlm,Isgur:1978xj,Isgur:1978wd,Isgur:1979be}.
It has been developed to address multiquark systems.
The detail of the QDCSM can be found in Refs.~\cite{Wu:1996fm,Huang:2011kf,ChenLZ,Ping:1998si,Wu:1998wu,Pang:2001xx,Ping:2000cb,Ping:2000dx,Ping:2008tp}.
In this section, we mainly introduce the salient features of this model.
The general form of the pentaquark Hamiltonian is given by
	\begin{align}
		H=&\sum_{i=1}^5\left(m_i+\frac{\boldsymbol{p}_{i}^{2}}{2m_i}\right)-T_{\mathrm{c} . \mathrm{m}} +\sum_{j>i=1}^5 V(\boldsymbol{r}_{ij}),
	\end{align}
where $m_i$ is the quark mass, $\boldsymbol{p}_{i}$ is the momentum of the quark, and $T_{\mathrm{c.m.}}$ is the center-of-mass kinetic energy.
The dynamics of the pentaquark system is driven by a two-body potential
	\begin{align}
		V(\boldsymbol{r}_{ij})= & V_{\mathrm{CON}}(\boldsymbol{r}_{ij})+V_{\mathrm{OGE}}(\boldsymbol{r}_{ij})+V_{\chi}(\boldsymbol{r}_{ij}).
	\end{align}
The most relevant features of QCD at its low energy regime---color confinement ($V_{\mathrm{CON}}$), perturbative one-gluon exchange interaction ($V_{\mathrm{OGE}}$), and dynamical chiral symmetry breaking ($V_{\chi}$)---have been taken into consideration.

Here, a phenomenological color screening confinement potential ($V_{\mathrm{CON}}$) is used as
\begin{align}
	V_{\mathrm{CON}}(\boldsymbol{r}_{ij}) = & -a_{c}\boldsymbol{\lambda}_{i}^{c} \cdot \boldsymbol{\lambda}_{j}^{c}\left[  f(\boldsymbol{r}_{ij})+V_{0}\right],
\end{align}
\begin{align}
	f(\boldsymbol{r}_{ij}) =& \left\{\begin{array}{l}
		\boldsymbol{r}_{i j}^{2}, ~~~~~~~~~~~~~ ~i,j ~\text {occur in the same cluster } \\
		\frac{1-e^{-\mu_{q_{i}q_{j}} \boldsymbol{r}_{i j}^{2}}}{\mu_{q_{i}q_{j}}},  ~~~i,j ~\text {occur in different cluster }
	\end{array}\right.   \nonumber
\end{align}
where $a_c$, $V_{0}$ and $\mu_{q_{i}q_{j}}$ are model parameters, and $\boldsymbol{\lambda}^{c}$ stands for the SU(3) color Gell-Mann matrices.
Among them, the color screening parameter $\mu_{q_{i}q_{j}}$ is determined by fitting the deuteron properties, nucleon-nucleon scattering phase shifts, and hyperon-nucleon scattering phase shifts, respectively, with $\mu_{qq}=0.45$, $\mu_{qs}=0.19$, and $\mu_{ss}=0.08~$fm$^{-2}$, satisfying the relation---$\mu_{qs}^{2}=\mu_{qq}\mu_{ss}$~\cite{ChenM}.
Additionally, we found that the heavier the quark, the smaller this parameter $\mu_{q_{i}q_{j}}$.
When extending to the heavy-quark system, the hidden-charm pentaquark system, we took $\mu_{cc}$ as an adjustable parameter from 0.01 to $0.001~$fm$^{-2}$, and found that the results were insensitive to the value of $\mu_{cc}$~\cite{HuangPc1}.
Moreover, the $P_{c}$ states were well predicted in the work of Refs.~\cite{HuangPc1,HuangPc2}.
So here we take $\mu_{cc}=0.01$ and $\mu_{qc}=0.067~$fm$^{-2}$, also satisfying the relation---$\mu_{qc}^{2}=\mu_{qq}\mu_{qc}$.

In the present work, we mainly focus on the low-lying negative parity $ssc\bar{q}q$ pentaquark states of the $S$-wave, so the spin-orbit and tensor interactions are not included.
The one-gluon exchange potential ($V_{\mathrm{OGE}}$), which includes Coulomb and chromomagnetic interactions, is written as
	\begin{align}
		V_{\mathrm{OGE}}(\boldsymbol{r}_{ij})= &\frac{1}{4}\alpha_{s_{q_i q_j}} \boldsymbol{\lambda}_{i}^{c} \cdot \boldsymbol{\lambda}_{j}^{c}  \\
		&\cdot \left[\frac{1}{r_{i j}}-\frac{\pi}{2} \delta\left(\mathbf{r}_{i j}\right)\left(\frac{1}{m_{i}^{2}}+\frac{1}{m_{j}^{2}}+\frac{4 \boldsymbol{\sigma}_{i} \cdot \boldsymbol{\sigma}_{j}}{3 m_{i} m_{j}}\right)\right],   \nonumber \label{Voge}
	\end{align}
where $\boldsymbol{\sigma}$ is the Pauli matrices and $\alpha_{s_{q_i q_j}}$ is the quark-gluon coupling constant.

However, the quark-gluon coupling constant between quark and anti-quark, which offers a consistent description of mesons from light to heavy-quark sector, is determined by the mass differences between pseudoscalar mesons (spin-parity $J^P=0^-$) and vector (spin-parity $J^P=1^-$), respectively.
For example, from the model Hamiltonian, the mass difference between $\bar{D}$ and $\bar{D}^*$ is determined by the chromomagnetic interaction in Eq.~(\ref{Voge}), so the parameter $\alpha_{s_{\bar{c}q}}$ is determined by fitting the mass difference between $\bar{D}$ and $\bar{D}^*$.

The dynamical breaking of chiral symmetry results in the SU(3) Goldstone boson exchange interactions appear between constituent light quarks $u, d$, and $s$.
Hence, the chiral interaction is expressed as
\begin{align}
	V_{\chi}(\boldsymbol{r}_{ij})= & V_{\pi}(\boldsymbol{r}_{ij})+V_{K}(\boldsymbol{r}_{ij})+V_{\eta}(\boldsymbol{r}_{ij}).
\end{align}
Among them
\begin{align}
V_{\pi}\left(\boldsymbol{r}_{i j}\right) =&\frac{g_{c h}^{2}}{4 \pi} \frac{m_{\pi}^{2}}{12 m_{i} m_{j}} \frac{\Lambda_{\pi}^{2}}{\Lambda_{\pi}^{2}-m_{\pi}^{2}} m_{\pi}\left[Y\left(m_{\pi} \boldsymbol{r}_{i j}\right)\right. \nonumber \\
&\left.-\frac{\Lambda_{\pi}^{3}}{m_{\pi}^{3}} Y\left(\Lambda_{\pi} \boldsymbol{r}_{i j}\right)\right]\left(\boldsymbol{\sigma}_{i} \cdot \boldsymbol{\sigma}_{j}\right) \sum_{a=1}^{3}\left(\boldsymbol{\lambda}_{i}^{a} \cdot \boldsymbol{\lambda}_{j}^{a}\right),
\end{align}
\begin{align}
	V_{K}\left(\boldsymbol{r}_{i j}\right) =&\frac{g_{c h}^{2}}{4 \pi} \frac{m_{K}^{2}}{12 m_{i} m_{j}} \frac{\Lambda_{K}^{2}}{\Lambda_{K}^{2}-m_{K}^{2}} m_{K}\left[Y\left(m_{K} \boldsymbol{r}_{i j}\right)\right. \nonumber \\
	&\left.-\frac{\Lambda_{K}^{3}}{m_{K}^{3}} Y\left(\Lambda_{K} \boldsymbol{r}_{i j}\right)\right]\left(\boldsymbol{\sigma}_{i} \cdot \boldsymbol{\sigma}_{j}\right) \sum_{a=4}^{7}\left(\boldsymbol{\lambda}_{i}^{a} \cdot \boldsymbol{\lambda}_{j}^{a}\right),
\end{align}
\begin{align}
	V_{\eta}\left(\boldsymbol{r}_{i j}\right) =&\frac{g_{c h}^{2}}{4 \pi} \frac{m_{\eta}^{2}}{12 m_{i} m_{j}} \frac{\Lambda_{\eta}^{2}}{\Lambda_{\eta}^{2}-m_{\eta}^{2}} m_{\eta}\left[Y\left(m_{\eta} \boldsymbol{r}_{i j}\right)\right. \nonumber \\
	&\left.-\frac{\Lambda_{\eta}^{3}}{m_{\eta}^{3}} Y\left(\Lambda_{\eta} \boldsymbol{r}_{i j}\right)\right]\left(\boldsymbol{\sigma}_{i} \cdot \boldsymbol{\sigma}_{j}\right)\left[\cos \theta_{p}\left(\boldsymbol{\lambda}_{i}^{8} \cdot \boldsymbol{\lambda}_{j}^{8}\right)\right.  \nonumber \\
	&\left.-\sin \theta_{p}\left(\boldsymbol{\lambda}_{i}^{0} \cdot \boldsymbol{\lambda}_{j}^{0}\right)\right],
\end{align}
where $Y(x) = e^{-x}/x$ is the standard Yukawa function.
The physical $\eta$ meson is considered by introducing the angle $\theta_{p}$ instead of the octet one. The $\boldsymbol{\lambda}^a$ are the SU(3) flavor Gell-Mann matrices.
The values of $m_\pi$, $m_k$ and $m_\eta$ are the masses of the SU(3) Goldstone bosons, which adopt the experimental values~\cite{Workman:2022ynf}.
The chiral coupling constant $g_{ch}$, is determined from the $\pi N N$ coupling constant through
\begin{align}
	\frac{g_{c h}^{2}}{4 \pi} & = \left(\frac{3}{5}\right)^{2} \frac{g_{\pi N N}^{2}}{4 \pi} \frac{m_{u, d}^{2}}{m_{N}^{2}}.
\end{align}
Assuming that flavor SU(3) is an exact symmetry, it will only be broken by the different mass of the strange quark.
The other symbols in the above expressions have their usual meanings.
All the parameters shown in Table~\ref{parameters} are fixed by masses of the ground-state baryons and mesons.
Table~\ref{hadrons} shows the masses of the baryons and mesons used in this work.
Since it is very difficult to fit well all ground-state hadrons with limited parameters, we give priority to fitting lighter baryons and mesons when setting parameters.
As a result, the mass gaps between theoretical and experimental values of heavier baryons and mesons are larger.

In the QDCSM, quark delocalization was introduced to enlarge the model variational space to take into account the mutual distortion or the internal excitations of nucleons in the course of interaction.
It is realized by specifying the single-particle orbital wave function of the QDCSM as a linear combination of left and right Gaussians, the single-particle orbital wave functions used in the ordinary quark cluster model
\begin{eqnarray}
	\psi_{\alpha}(\boldsymbol {S_{i}} ,\epsilon) & = & \left(
	\phi_{\alpha}(\boldsymbol {S_{i}})
	+ \epsilon \phi_{\alpha}(-\boldsymbol {S_{i}})\right) /N(\epsilon), \nonumber \\
	\psi_{\beta}(-\boldsymbol {S_{i}} ,\epsilon) & = &
	\left(\phi_{\beta}(-\boldsymbol {S_{i}})
	+ \epsilon \phi_{\beta}(\boldsymbol {S_{i}})\right) /N(\epsilon), \nonumber \\
	N(S_{i},\epsilon) & = & \sqrt{1+\epsilon^2+2\epsilon e^{-S_i^2/4b^2}}. \label{1q}
\end{eqnarray}
It is worth noting that the mixing parameter $\epsilon$ is not an adjusted one but determined variationally by the dynamics of the multiquark system itself.
In this way, the multiquark system chooses its favorable configuration in the interacting process.
This mechanism has been used to explain the crossover transition between the hadron phase and quark-gluon plasma phase~\cite{Xu}.

In addition, the dynamical calculation is carried out using the resonating group method and the generating coordinates method.
The details of the two methods are presented in Appendix, and the way of constructing wave functions can be seen in Ref.~\cite{Xia:2021tof}.
\begin{table}[ht]
	\caption{\label{parameters}Model parameters used in this work:
		$m_{\pi} = 0.7$, $m_{K} = 2.51$, $m_{\eta} = 2.77$, $\Lambda_{\pi} = 4.2$, $\Lambda_{K} = 5.2$, $\Lambda_{\eta} = 5.2$ fm$^{-1}$, $g_{ch}^2/(4\pi)$ = 0.54.}
	\begin{tabular}{cccccc} \hline\hline
		~~~~$b$~~~~ & ~~$m_{q}$~~ & ~~~$m_{c}$~~~  & ~~~$V_{0_{qq}}$~~~~&~~~$V_{0_{q\bar{q}}}$~~~~& ~~~$ a_c$~~~   \\
		(fm)        & (MeV)       & (MeV)          & (fm$^{-2}$)        & (fm$^{-2}$)             & ~(MeV\,fm$^{-2}$)~ \\
		0.518       & 313         & 1788           &  $-$1.288          &  0.207                  &  58.03 \\ \hline
		$\alpha_{s_{\bar{c}q}}$ & $\alpha_{s_{\bar{c}c}}$ &$\alpha_{s_{qq}}$   & $\alpha_{s_{qc}}$ &$\alpha_{s_{cc}}$   \\
		1.970                   &  2.357  &    0.524                  & 0.467             &   0.213              \\ \hline\hline	
	\end{tabular}
\end{table}

\begin{table}[ht]
	\caption{The masses (in MeV) of the baryons and mesons. Experimental values are taken from the Particle Data Group~\cite{Workman:2022ynf}.}
	\begin{tabular}{c c c c}
		\hline \hline
		Hadron & ~~~~~$I(J^P)$~~~~~  & ~~~~~~$M^{\text{Exp}}$~~~~~~ & ~$M^{\text{Theo}}$~  \\ \hline
		$\bar{D}$        & $1/2(0^-)$   & 1869  & 1869   \\
        $\bar{D}^*$      & $1/2(1^-)$   & 2007  & 2007   \\
        $\eta_c$         & $0(0^-)$     & 2984  & 2984   \\
        $J/\psi$         & $0(1^-)$     & 3097  & 3013   \\
		$N$              & $1/2(1/2^+)$ & 939   & 939    \\
		$\Delta$         & $3/2(3/2^+)$ & 1232  & 1232   \\
		$\Lambda_c$      & $0(1/2^+)$   & 2286  & 2286   \\
		$\Sigma_c$       & $1(1/2^+)$   & 2455  & 2465   \\
		$\Sigma^*_c$     & $1(3/2^+)$   & 2490  & 2518   \\
		$\Xi_{cc}$       & $1/2(1/2^+)$ & 3621  & 3766   \\
		$\Xi_{cc}^*$     & $1/2(3/2^+)$ &       & 3791   \\

		 \hline\hline
	\end{tabular}
	\label{hadrons}
\end{table}

\section{The results and discussions}
In the present calculation, we systematically investigate the $S$-wave $qqc\bar{c}c$ ($q=u~\text{or}~d)$ pentaquark systems in the framework of the QDCSM.
The quantum numbers $I$ = 0 and 1, $J^P = 1/2^-, 3/2^-$, and $5/2^-$ are considered.
First, we study the effective potential of each channel, which is presented in the Fig.~\ref{I=0} and Fig.~\ref{I=1}.
Moreover, to find out if there exists any bound state, we carry out a dynamic bound-state calculation of both single-channel and channel coupling.
The root mean square (rms) of cluster spacing of the obtained state is calculated to determine the spatial configuration.
Additionally, the scattering process is also studied to search for the resonance state, and the detail of this process is introduced in Appendix.
In addition, we further investigate the influence of different interaction terms in the effective potential and their contribution to the binding energy, to explore the nature of the obtained states.

\subsection{Effective potential}

\begin{figure*}[htbp]
	\centering
	\includegraphics[width=18cm]{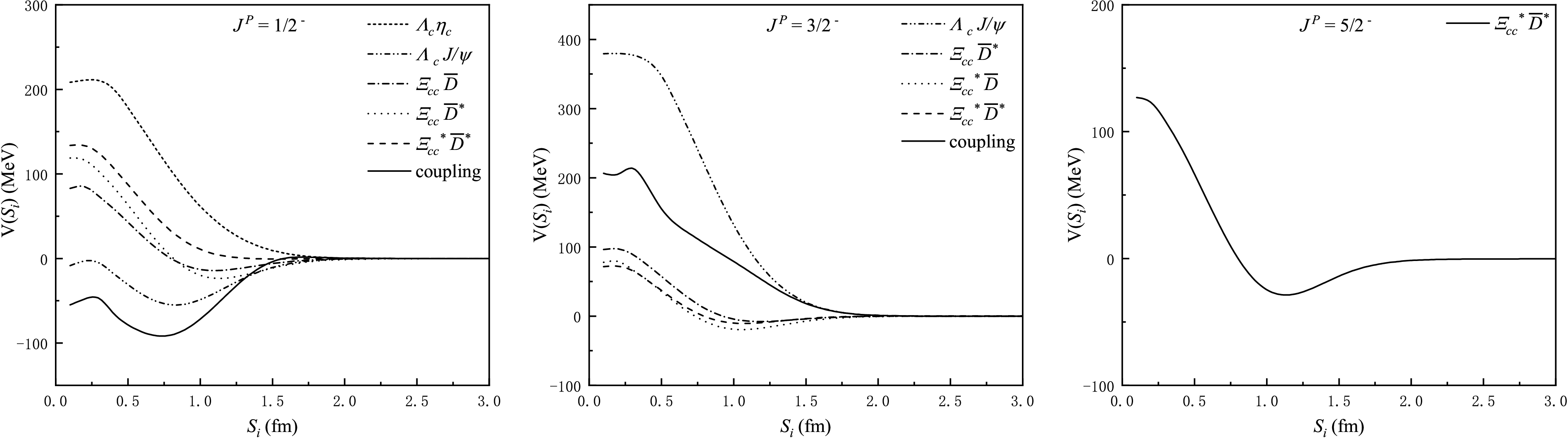}\
	\caption{\label{I=0} The effective potential of $qqc\bar{c}c$ system with $I=0$.}
\end{figure*}

\begin{figure*}[htbp]
	\centering
	\includegraphics[width=18cm]{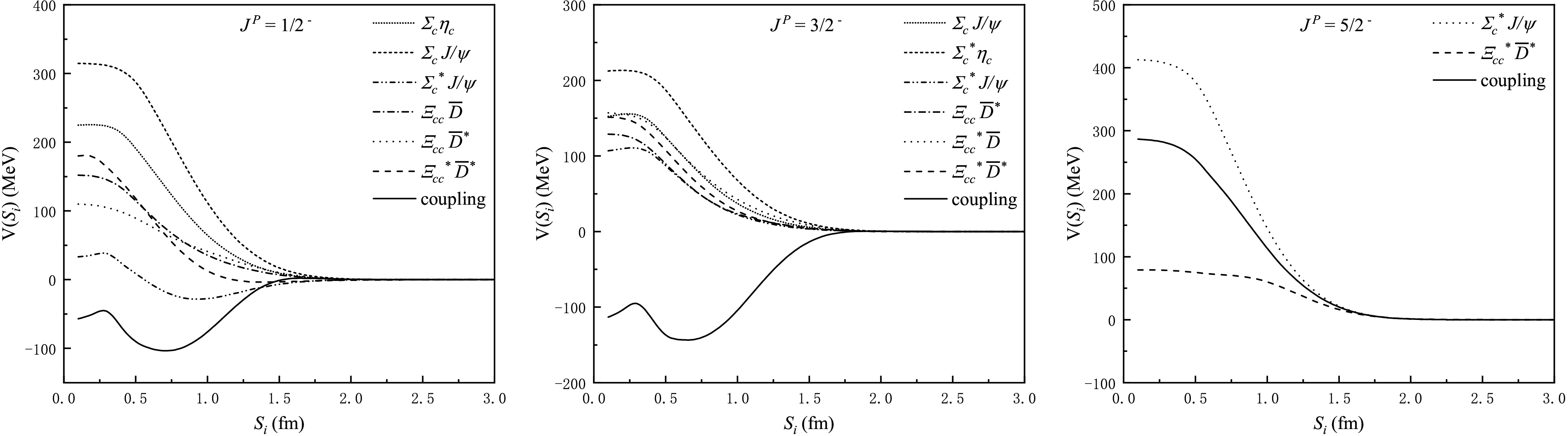}\
	\caption{\label{I=1} The effective potential of $qqc\bar{c}c$ system with $I=1$.}
\end{figure*}

Since an attractive potential is necessary for forming a bound state, we first calculate the effective potential.
It is defined as $V(S_i) = E(S_i) - E(\infty)$, where $S_{i}$ stands for the distance between two clusters.
$E(S_i)$ and $E(\infty)$ are the energies of the system at the generator coordinate $S_{i}$ and at a sufficient large distance, respectively.
$E(S_i)$ is obtained as:
\begin{align}
	E\left(S_{i}\right) & = \frac{\left\langle\Psi_{5 q}\left(S_{i}\right)|H| \Psi_{5 q}\left(S_{i}\right)\right\rangle}{\left\langle\Psi_{5 q}\left(S_{i}\right) \mid \Psi_{5 q}\left(S_{i}\right)\right\rangle},
\end{align}
where $\Psi_{5 q}(S_{i})$ represents the wave function of a certain channel, $\left\langle\Psi_{5 q}\left(S_{i}\right)|H| \Psi_{5 q}\left(S_{i}\right)\right\rangle$ and $\left\langle\Psi_{5 q}\left(S_{i}\right) \mid \Psi_{5 q}\left(S_{i}\right)\right\rangle$ are the diagonal matrix element of the Hamiltonian and the overlap of the system.
In order to investigate the interaction between baryons and mesons, the effective potential of each channel with different quantum numbers is presented in Fig.~\ref{I=0} and~\ref{I=1}.
In addition, after channel coupling, the lowest energies are chosen to draw the effective potential of coupled-channel.

For the $I({J^P}) = 0(1/2^-)$ system, in Fig.~\ref{I=0},  there are five physical channels: the $\Lambda_c \eta_c$, $\Lambda_c J/\psi$, $\Xi_{cc} \bar{D}$, $\Xi_{cc} \bar{D}^*$, and $\Xi_{cc}^* \bar{D}^*$.
One can see that the effective potentials of the $\Lambda_c \eta_c$ and $\Xi_{cc}^* \bar{D}^*$ channels are purely repulsive.
Therefore, the $\Lambda_c \eta_c$ and $\Xi_{cc}^* \bar{D}^*$ cannot form bound state in the single channel calculation due to the lack of attraction.
Additionally, the effective potentials of the $\Xi_{cc} \bar{D}$ and $\Xi_{cc} \bar{D}^*$ show weakly attraction at about 1.2 fm, but have relatively large repulsion at close range.
Quantum mechanics tells us that the distance between two clusters is probabilistic from close range to large range, so both interactions at close range and medium range play a role in the formation of state.
In this case, the $\Xi_{cc} \bar{D}$ and $\Xi_{cc} \bar{D}^*$ are difficult to form bound state.
As for the $\Lambda_c J/\psi$, the effective potential at medium range is attractive and there is no repulsion at close range.
It is likely for the $\Lambda_c J/\psi$ to form a bound state and a dynamic calculation is carried out in the next subsection.
After channel coupling, there is a strong attraction, indicating that the coupled-channel is likely to be bound.

For the $I({J^P}) = 0(3/2^-)$ system, the $\Xi_{cc} \bar{D}^*$, $\Xi_{cc}^* \bar{D}$, and $\Xi_{cc}^* \bar{D}^*$ show very weak attraction, which is not enough to form a bound state.
The effective potentials of the $\Lambda_c \eta_c$ and coupled-channel are all purely repulsive.
Besides, one might wonder why the effective potential of the coupled-channel is not the lowest one.
This is because effective potential is obtained as $V(S_i) = E(S_i) - E(\infty)$, where $E(\infty)$ of different channel is different.
Although the $E(S_i)$ of the coupled-channel the lower than that of every single channel, the $E(\infty)$ of the coupled-channel is also the lowest one.
Therefore, sometimes the effective potential of the coupled-channel is not the lowest one.
Moreover, for the $I({J^P}) = 0(3/2^-)$ system, the main composition of the coupled-channel is the $\Lambda_c J/\psi$.
Compared with the $\Lambda_c J/\psi$, the repulsion of the coupled-channel is not as strong, but it still cannot form a bound state.

Since our current calculation is based on the $S$-wave system ($L = 0$), there is only $\Xi_{cc}^* \bar{D}^*$ can be coupled to a system with $I({J^P}) = 0(5/2^-)$.
And there is no need for channel coupling.
One can find that the effective potential is attractive at medium range, indicating that it is possible to form a bound state.

For the $I({J^P}) = 1(1/2^-)$ system, there are six channels, which are the $\Sigma_c \eta_c$, $\Sigma_c J/\psi$, $\Sigma_c^* J/\psi$, $\Xi_{cc} \bar{D}$, $\Xi_{cc} \bar{D}^*$, and $\Xi_{cc}^* \bar{D}^*$.
Only the $\Sigma_c^* J/\psi$ shows a weak attraction, and the repulsion at close range is also very weak, making it still possible to be a bound state.
After channel coupling, the attraction of coupled-channel is deepened and there is almost no repulsive interaction.
It is likely for the $qqc\bar{c}c$ with $I({J^P}) = 1(1/2^-)$ to form bound state.

For the $I({J^P}) = 1(3/2^-)$ system, the effective potential of all single-channels are repulsive.
However, the coupled-channel show an obvious attraction, which may result in a bound state.
As for the $I({J^P}) = 1(5/2^-)$ system, there are two channels the $\Sigma_c^* J/\psi$ and $\Xi_{cc}^* \bar{D}^*$.
The two channels together with the coupled-channel are all purely repulsive.
Therefore, the the $qqc\bar{c}c$ with $I({J^P}) = 1(5/2^-)$ cannot form any bound state.

\subsection{Bound-state calculation}

In order to confirm whether any state with attractive interaction can form a bound state, a dynamic calculation is carried out in this part.
The numerical results of pentaquark systems with different quantum numbers are listed in Table~\ref{1}.
The first column headed with $I(J^P)$ is the quantum number of each system.
The second column headed with "Channel", gives the physical channels involved in the present work.
The third column headed with $E_{\mathrm{th}}$ refers to the theoretical value of non-interacting baryon-meson threshold.
The forth column headed with $E_{\mathrm{sc}}$ shows the energy of each single channel.
The values of binding energy $E_\mathrm{B}$= $E_{\mathrm{sc}} -E_{\mathrm{th}}^{\mathrm{Theo}}$ are listed in the fifth column only if $E_\mathrm{B} < 0$ MeV.
Otherwise there will be "Ub", which represents that the system is unbound.
The experimental threshold $E_{\mathrm{th}}^{\mathrm{Exp}}$, which is the sum of the experimental masses of the corresponding baryon and meson, and the corrected mass $E^{\prime} = E_{\mathrm{th}}^{\mathrm{Exp}} + E_\mathrm{B}$ are given in the sixth and seventh columns.
After channel coupling, the lowest energy of coupled-channel $E_{\mathrm{cc}}$ and the corresponding binding energy $E_\mathrm{B}$ is listed in the eighth column.
Finally, the corrected mass of coupled-channel $E^\prime_{\mathrm{cc}}$ is given in the last column.
The definitions of $E_{\mathrm{cc}}$ and $E_\mathrm{B}$ in channel coupling calculation are similar to that in single-channel calculation.
When we deal with the mass correction of the coupled-channel, a modified formula~\cite{Hu:2022pae,Yan:2022nxp} is used---$E^{\prime}_{\mathrm{cc}}=E_{\mathrm{cc}}+\sum_{n} p_{n}\left[E_{\mathrm{th}}^{\mathrm{Exp}}(n)-E_{\mathrm{th}}^{\mathrm{Theo}}(n)\right]$, where $p_{n}$ is the proportion of $n$th physical channel.
In addition, the experimental value of $\Xi_{cc}^*$ has not been measured so far.
Here we use the calculated value of $\Xi_{cc}^*$ in Ref.~\cite{Ortiz-Pacheco:2023kjn} for mass correction.

It is worth noting that, only the lowest energy of each channel is presented in the table, because whether the system can form bound state depends on whether the lowest energy is below the threshold.
First of all, from Table~\ref{1}, an intuitive analysis can be based on the results of the single-channel calculation.
For the $I({J^P}) = 0(1/2^-)$ system, only the $\Lambda_c J/\psi$ forms bound state in the single-channel calculation, and other four channels are all scattering states.
The binding energy of the $\Lambda_c J/\psi$ is about $-$15 MeV.
However, the $\Lambda_c J/\psi$ can decay to $\Lambda_c \eta_c$ through strong interaction.
In order to determine whether the $\Lambda_c J/\psi$ can be a resonance state, the scattering phase shift of this process is studied in the next part.
Although the $\Xi_{cc} \bar{D}$ and $\Xi_{cc} \bar{D}^*$ have attractive effective potentials, it turns out that the weak attractions are not enough to form any bound state.
According to the effective potential of coupled-channel, the interaction between baryon and meson is attractive.
The numerical results also show that the channel coupling forms a bound state which is 49 MeV below the lowest threshold of the $I({J^P}) = 0(1/2^-)$ system.
The main composition of this state is the $\Lambda_c \eta_c$ and $\Lambda_c J/\psi$ according to the numerical calculation.
Since in the current work we only consider $S$-wave systems, this bound state cannot to other channels.
Moreover, we carry out the mass correction and the rms calculation of this bound state.
The corrected mass of the obtained state is 5259 MeV and the rms is 1.0 fm.
On the basis of its relatively small rms, it is a compact pentaquark state.

\begin{table*}[htb]
	\caption{\label{1} The energies of single-channels and coupled-channels (in MeV).}
	\begin{tabular}{c c c c c c c c c} \hline\hline
		
		~~~~$I(J^P)$~~~~ & ~~Channel~~ & ~~~~~~$E_{\mathrm{th}}^{\mathrm{Theo}}$~~~~~~ & ~~~~$E_{\mathrm{sc}}$~~~~ & ~~~~~$E_{\mathrm{B}}$~~~~~ &  ~~~~~$E_{\mathrm{th}}^{\mathrm{Exp}}$~~~~~ & ~~~~~$E_{\mathrm{sc}}^{\prime}$~~~~~ & ~~~~$E_{\mathrm{cc}}/E_{\mathrm{B}}$~~~~ & ~~~~$E_{\mathrm{cc}}^{\prime}$~~~~   \\ \hline
		
		$0(1/2^-)$ & $\Lambda_c \eta_c$     & 5232 & 5234 &  Ub  & 5270 & 5272 & 5183/$-$49 & 5259  \\
		           & $\Lambda_c J/\psi$     & 5261 & 5246 &$-$15 & 5383 & 5368 &            &       \\
		           & $\Xi_{cc} \bar{D}$     & 5644 & 5646 &  Ub  & 5490 & 5492 &            &       \\
		           & $\Xi_{cc} \bar{D}^*$   & 5782 & 5783 &  Ub  & 5628 & 5629 &            &       \\
		           & $\Xi_{cc}^* \bar{D}^*$ & 5807 & 5809 &  Ub  & 5693 & 5695 &            &       \\ \hline
		
		$0(3/2^-)$ & $\Lambda_c J/\psi$     & 5261 & 5363 &  Ub  & 5383 & 5385 & 5863/Ub    & 5385  \\
		           & $\Xi_{cc} \bar{D}^*$   & 5782 & 5784 &  Ub  & 5628 & 5630 &            &       \\
	               & $\Xi_{cc}^* \bar{D}$   & 5669 & 5670 &  Ub  & 5555 & 5556 &            &       \\
		           & $\Xi_{cc}^* \bar{D}^*$ & 5807 & 5809 &  Ub  & 5693 & 5695 &            &       \\\hline
		
		$0(5/2^-)$ & $\Xi_{cc}^* \bar{D}^*$ & 5807 & 5804 & $-$3 & 5693 & 5690 & 5804/$-$3  & 5690  \\ \hline
		
		$1(1/2^-)$ & $\Sigma_c \eta_c$      & 5411 & 5413 &  Ub  & 5439 & 5441 & 5341/$-$70 & 5396 \\
		           & $\Sigma_c J/\psi$      & 5439 & 5442 &  Ub  & 5552 & 5555 &            &       \\
		           & $\Sigma_c^*J/\psi$     & 5464 & 5463 & $-$2 & 5617 & 5616 &            &       \\
		           & $\Xi_{cc} \bar{D}$     & 5644 & 5647 &  Ub  & 5490 & 5493 &            &       \\
		           & $\Xi_{cc} \bar{D}^*$   & 5782 & 5785 &  Ub  & 5628 & 5631 &            &       \\
		           & $\Xi_{cc}^* \bar{D}^*$ & 5807 & 5809 &  Ub  & 5693 & 5695 &            &       \\ \hline
		
		$1(3/2^-)$ & $\Sigma_c J/\psi$      & 5439 & 5442 &  Ub  & 5552 & 5555 & 5357/$-$78 & 5465  \\
		           & $\Sigma_c^*\eta_c$     & 5435 & 5437 &  Ub  & 5504 & 5506 &            &       \\
		           & $\Sigma_c^*J/\psi$     & 5464 & 5466 &  Ub  & 5617 & 5619 &            &       \\
		           & $\Xi_{cc} \bar{D}^*$   & 5782 & 5785 &  Ub  & 5628 & 5631 &            &       \\
		           & $\Xi_{cc}^* \bar{D}$   & 5669 & 5671 &  Ub  & 5555 & 5557 &            &       \\
		           & $\Xi_{cc}^* \bar{D}^*$ & 5807 & 5809 &  Ub  & 5693 & 5695 &            &       \\\hline
		
		$1(5/2^-)$ & $\Sigma_c^*J/\psi$     & 5464 & 5466 &  Ub  & 5617 & 5619 & 5466/Ub    & 5619  \\
		           & $\Xi_{cc}^* \bar{D}^*$ & 5807 & 5809 &  Ub  & 5693 & 5695 &            &       \\ \hline\hline
	\end{tabular}
\end{table*}

For the $I({J^P}) = 0(3/2^-)$ system, the energy of each single-channel is above the corresponding threshold.
Therefore, the four single-channels are all scattering states and cannot be resonance states.
After channel-coupling, the lowest energy of coupled-channel is still higher than the threshold of the lowest channel $\Lambda_c J/\psi$.
Based on this, there is no stable state in the $I({J^P}) = 0(3/2^-)$ system, which is consistent with the behavior of the corresponding effective potential.

For the $I({J^P}) = 0(5/2^-)$ system, a bound state $\Xi_{cc}^* \bar{D}^*$ with binding energy of $-$3 MeV is obtained.
The bound state $\Xi_{cc}^* \bar{D}^*$ can still decay to some $D$-wave channels, such as the $\Lambda_c J/\psi$, through the tensor force coupling.
However, we focus on the $S$-wave systems in this work, thus the tensor force coupling is not yet considered.
The $D$-wave decay will be the next step of our research in the future.
According to our previous research~\cite{Chen:2011zzb}, the decay width of $D$-wave decay is usually very narrow.
The corrected mass and the rms of this $\Xi_{cc}^* \bar{D}^*$ state is 5690 MeV and 1.9 fm, indicating that it is a molecular state.
Herein, an $I({J^P}) = 0(5/2^-)$ molecular state $\Xi_{cc}^* \bar{D}^*$ with energy of 5690 MeV is predicted.

For the $I({J^P}) = 1(1/2^-)$ system, the $\Sigma_c^* J/\psi$ forms a bound state with binding energy of $-$2 MeV in the single-channel calculation.
The other five channels turn out to be scattering states.
After channel coupling, the energy of the coupled-channel is 70 MeV lower than the lowest threshold, indicating the formation of a bound state.
Additionally, this state is dominated by the $\Sigma_c \eta_c$.
According to the further calculation, the corrected mass of this state is 5396 MeV and the rms is 0.9 fm.
Therefore, there is a possible compact pentaquark $qqc\bar{c}c$ with $I({J^P}) = 1(1/2^-)$, whose mass is 5396 MeV.

For the $I({J^P}) = 1(3/2^-)$ system, consistent with the behavior of effective potential, none of the single-channels is bound.
However, the effective potential shows that there is a strong attraction after channel coupling and the numerical result also confirms this.
The coupled-channel forms a deeply bound state, whose binding energy is $-78$ MeV.
This state is mainly composed of the $\Sigma_c J/\psi$ and $\Sigma_c^* \eta_c$ and its corrected mass is 5468 MeV.
As for the spatial configuration, it is determined to be a compact pentaquark state, based on its rms of 0.9 fm.

All energies of single-channels and coupled-channel of $I({J^P}) = 1(5/2^-)$ system are higher than the corresponding thresholds.
In other word, no stable state is found in this system.
The complete repulsion in the effective potential is also consistent with the numerical result.

\begin{figure}[htbp]
	\centering
	\includegraphics[width=8cm]{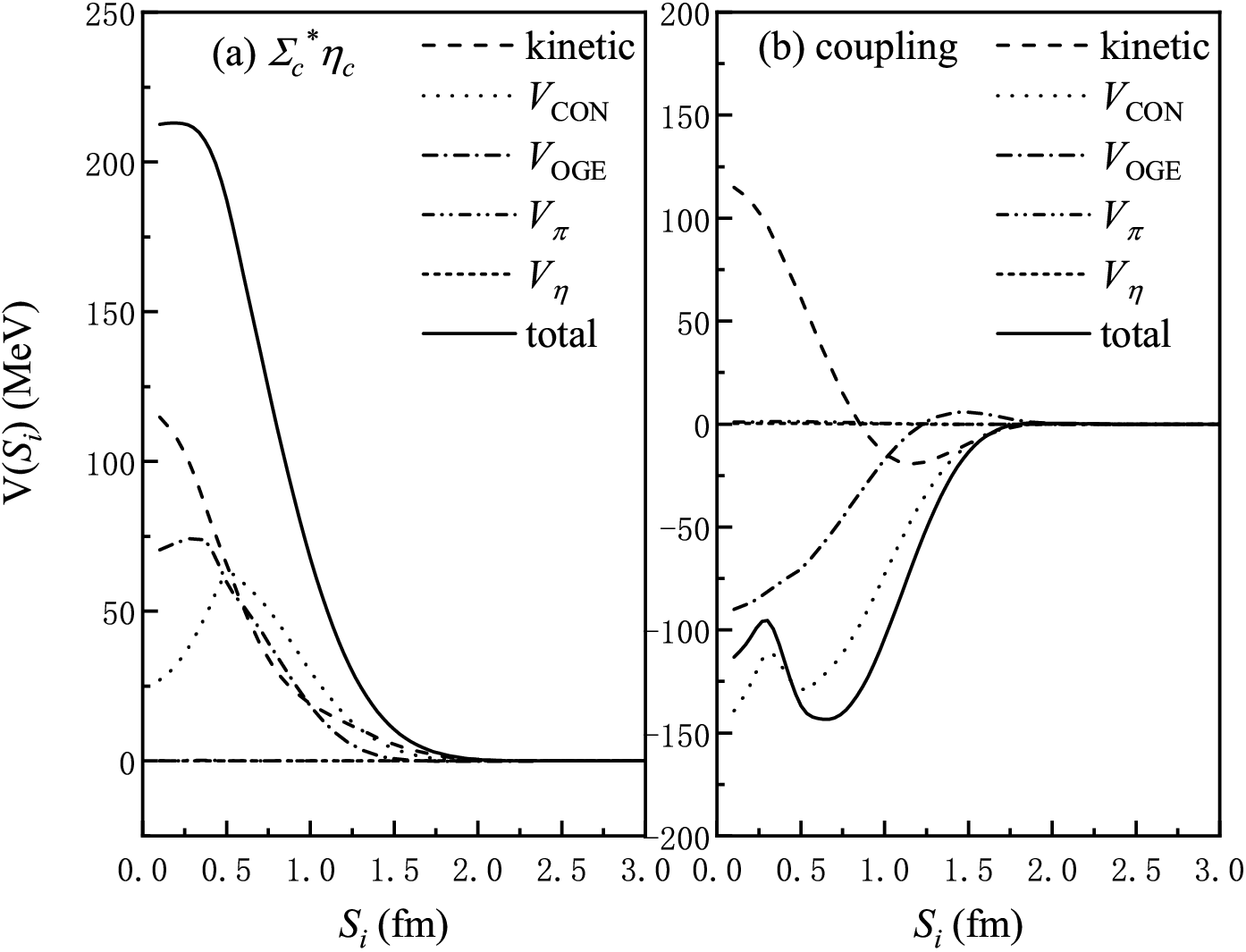}\
	\caption{\label{term} The effective potential of each interaction term of the $qqc\bar{c}c$ system with $I({J^P}) = 1(3/2^-)$, where (a) and (b) are the results of the single-channel and coupled-channel, respectively.}
\end{figure}

Further, one can find that none of the single-channels with $I({J^P}) = 1(3/2^-)$ are bound, but the coupled-channel is bound.
In order to figure out what causes this to happen, we investigate the effective potential of each interaction term.
Here we take the $qqc\bar{c}c$ with $I({J^P}) = 1(3/2^-)$ as an example.
The effective potential of each interaction term of the single-channel $\Sigma_c^* \eta_c$, which is the lowest energy threshold, and the coupled-channel is shown in Fig.~\ref{term}.
As we can see, the effect of $\pi$ exchange potential term ($V_\pi$) and $\eta$ exchange potential term ($V_\eta$) is very weak both before and after channel coupling.
The kinetic energy term always provides a large repulsive effect at the close range.
However, the repulsive effect is weakened after channel coupling.
The confinement potential term ($V_{\mathrm{CON}}$) and one-gluon exchange ($V_{\mathrm{OGE}}$) potential term are repulsive before channel coupling, but become obviously attractive after coupling.
Therefore, the mechanism of binding state formation is that channel coupling can weaken the repulsion of kinetic energy term, while confinement potential term and one-gluon exchange potential term provide strong attraction under the effect of channel coupling.
A similar situation also occurs in the $qqc\bar{c}c$ systems with $I({J^P}) = 0(1/2^-)$ and $I({J^P}) = 1(1/2^-)$.
Additionally, the contribution of each interaction term to the binding energy is calculated and listed in Table~\ref{contribution}.
For the $qqc\bar{c}c$ systems with $I({J^P}) = 0(1/2^-)$, $I({J^P}) = 1(1/2^-)$, and $I({J^P}) = 1(3/2^-)$, the repulsive contribution of kinetic energy term and the attractive contribution of $V_{\mathrm{CON}}$ and $V_{\mathrm{OGE}}$ are consistent with the result of effective potential.

\begin{table}[htb]
	\caption{\label{contribution} The contribution of each interaction term to the binding energy after channel coupling (in MeV).}
	\begin{tabular}{c c c c c c c} \hline\hline
		
		~$I(J^P)$~ & mass & ~kinetic~ & ~$V_{\text{CON}}$~ & ~~$V_{\text{OGE}}$~~ & ~~$V_{\pi, \eta}$~~ &  ~~total~~  \\ \hline
		
		$0(1/2^-)$ & 0    & +64.1     & $-$71.8     & $-$41.4       & +0.0           & $-$49.1  \\
		$1(1/2^-)$ & 0    & +75.1     & $-$126.4    & $-$20.3       & +1.5           & $-$70.0  \\
		$1(3/2^-)$ & 0    & +65.5     & $-$92.1     & $-$52.8       & +1.0           & $-$78.4   \\ \hline\hline
	\end{tabular}
\end{table}

\begin{figure}[htbp]
	\centering
	\includegraphics[width=8cm]{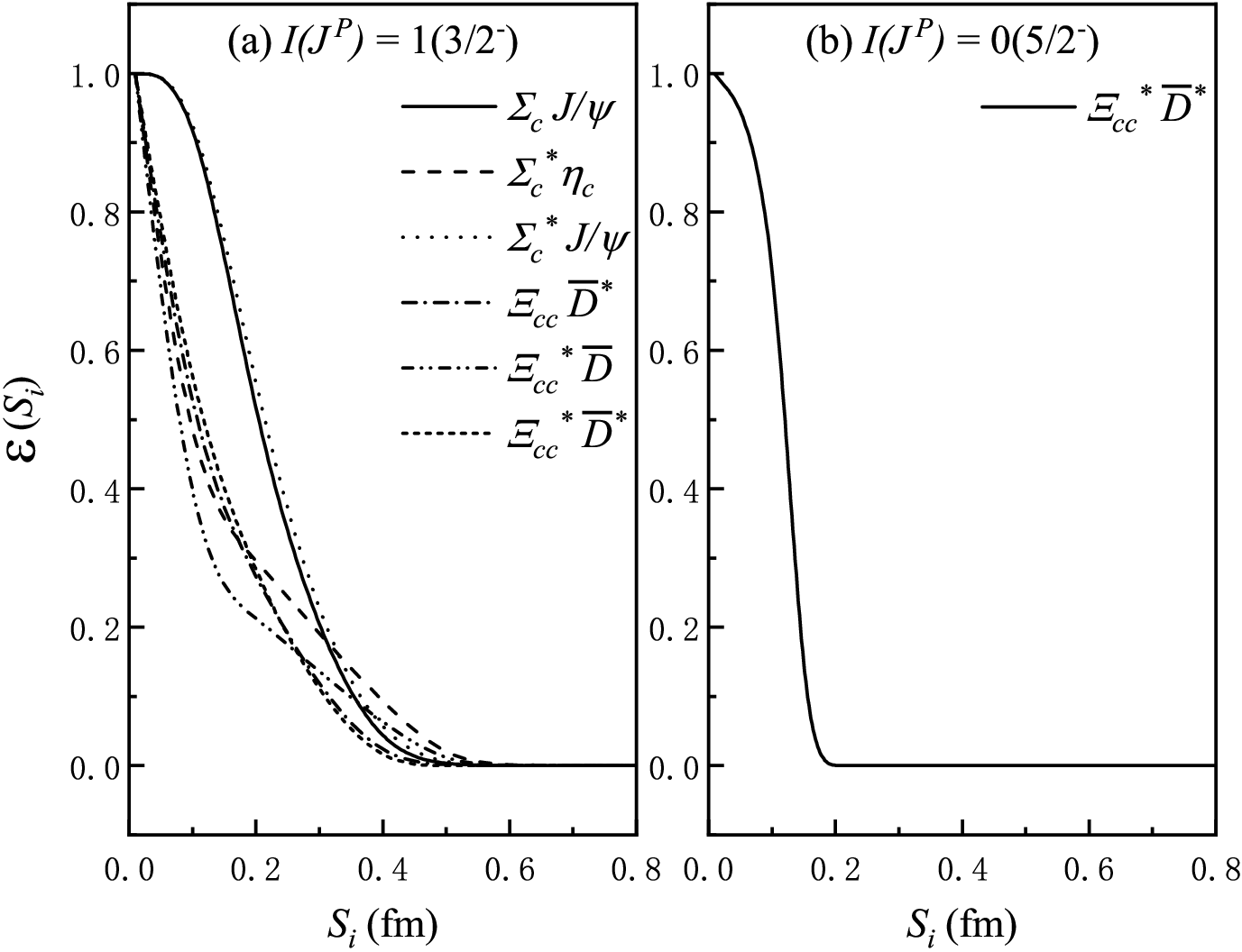}\
	\caption{\label{eps} The variation of delocalization parameter $\epsilon$ of each channel with (a) $I({J^P}) = 1(3/2^-)$ and (b) $I({J^P}) = 0(5/2^-)$.}
\end{figure}

Another factor that plays an important role in channel coupling is the delocalization effect of the model.
This is one of the features of our model, which allows quarks to run between baryon and meson.
The delocalization parameter $\epsilon$ is determined variationally by the dynamics of the system itself.
Thus, the pentaquark system chooses its favorable configuration in the interacting process.
According to this, the variational space of the system is expanded and the strength of interaction between two clusters (the baryon cluster and meson cluster) can be reflected by the delocalization parameter $\epsilon$.
Here we continue to take the $I({J^P}) = 1(3/2^-)$ system as an example, the delocalization parameter $\epsilon$ of each channel is shown in Fig.~\ref{eps} (a).
First, at extremely close range ($S_i < 0.1$ fm), the delocalization parameter $\epsilon$ approaches 1 because the baryon and meson can no longer be divided into two clusters when they are very close.
However, based on the value of $\epsilon$, the role of delocalization effect in the $I({J^P}) = 1(3/2^-)$ system is still evident at close range (0.1 $<S_i<$ 0.6 fm).
The increase of the interaction strength between baryon and meson caused by the delocalization effect also leads to the enhancement of the channel coupling effect.
Therefore, the repulsion of effective potential of single-channel becomes attractive after channel coupling at close range.

On the other hand, the trend of change in the delocalization parameter $\epsilon$ can also indirectly reflect the possible structure of the system.
Due to the influence of delocalization, there is mixing between the baryon and meson clusters of the $I({J^P}) = 1(3/2^-)$ system.
Although the composition calculation shows that the main components of the $I({J^P}) = 1(3/2^-)$ system are $\Sigma_c J/\psi$ and $\Sigma_c^* \eta_c$, the real components are not pure $\Sigma_c J/\psi$ and $\Sigma_c^* \eta_c$ due to the delocalization effect.
The mixing between baryons and mesons makes it more of a compact pentaquark state.
This conclusion is consistent with the small value of rms = 0.9 fm, which also shows the characteristic of a compact structure.
As a comparison, the delocalization effect in the $I({J^P}) = 0(5/2^-)$ system is much smaller, which can be seen in Fig.~\ref{eps} (b).
The delocalization parameter has been approaching 0 since 0.2 fm.
In this case, this system retains its molecular structure.
The rms of this state is 1.9 fm, confirming that it is a molecular state.

\subsection{Scattering process and resonance states}

\begin{figure}[htb]
	\centering
	\includegraphics[width=8cm]{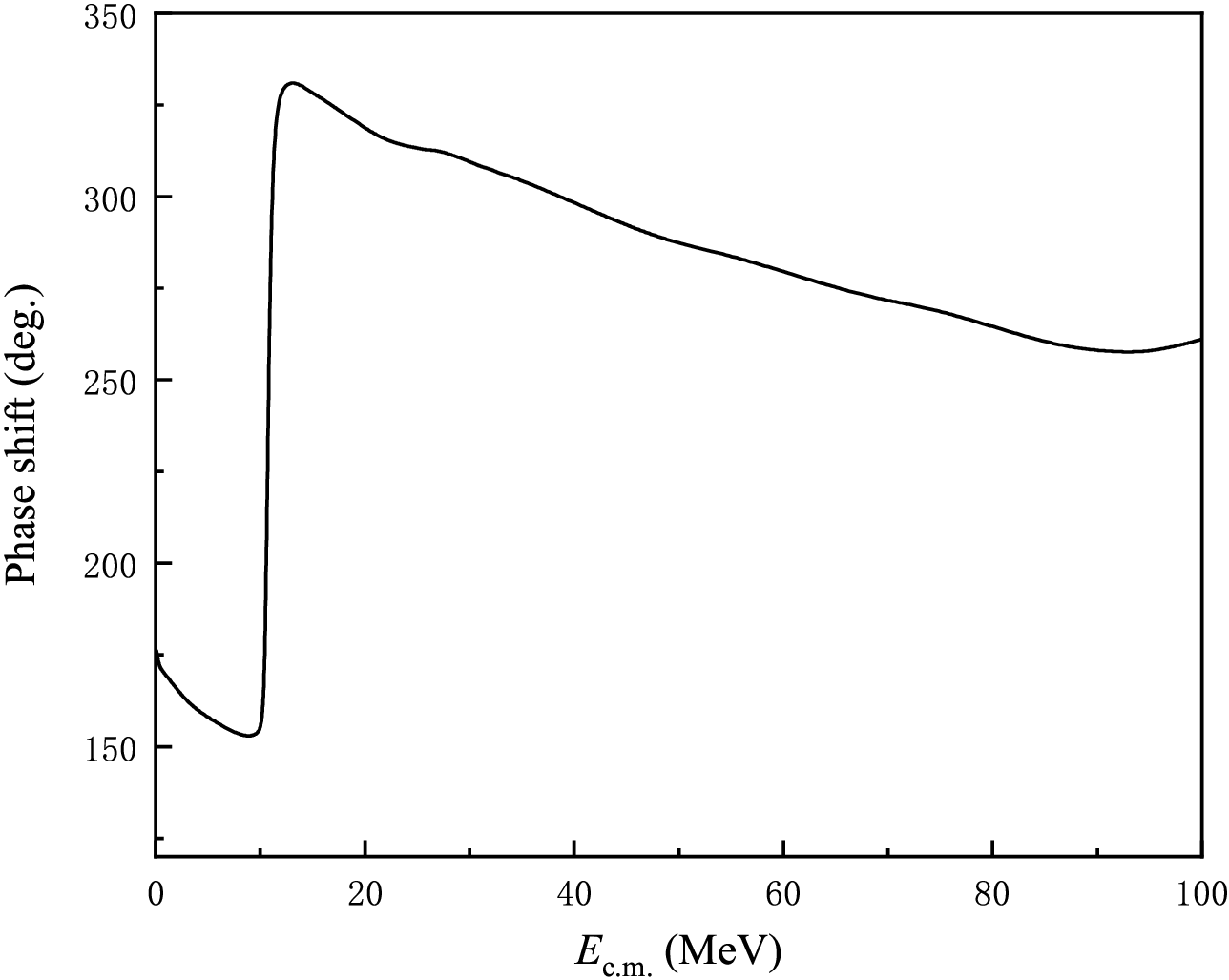}\
	\caption{\label{shift1}  The phase shift of the open channel $\Lambda_c \eta_c$ with $I(J^P) = 0(\frac{1}{2}^-)$ channel coupling.}
\end{figure}

As mentioned in the previous section, some quasibound states are obtained in the single-channel calculation, but the energy of each quasibound state is above the threshold of some channels of the corresponding system, which means that the quasibound states will decay to the lower channels and become resonance states or scattering states.
Therefore, the scattering phase shift calculation is performed to find resonance states.
First, the scattering phase shift of the $\Lambda_c \eta_c$ with $I({J^P}) = 0(1/2^-)$ is presented in Fig.~\ref{shift1} to test whether the $\Lambda_c J/\psi$ forms a resonance state.
A resonance state can usually be determined by the phase shift increased by 180$^\circ$.
The phase shift of open channel $\Lambda_c \eta_c$ shows 180$^\circ$ increase around the resonance mass, indicating that the $\Lambda_c J/\psi$ forms a resonance state.
Additionally, the way of identifying the mass and decay width of the resonance state through scattering phase shift can be seen in Appendix.
The resonance mass, corrected mass, decay width, and the value of the rms of this resonance state are summarized as follows:
\begin{align}
	M^{\mathrm{Theo}} &= 5245~ \mathrm{MeV},     \nonumber \\
	M^{\prime} &= 5367~ \mathrm{MeV},     \nonumber \\
	\Gamma &= 1~ \mathrm{MeV},      \nonumber \\
	\text{rms} &= 1.5~ \mathrm{fm}.      \nonumber
\end{align}
It is worth noting that, although the wave function of a resonance state is nonintegrable, we can calculate the rms of the main component of the resonance state, whose wave function is integrable.
According to the numerical result, the rms of the resonance state $\Lambda_c J/\psi$ is 1.5 fm, indicating that it is likely to be a molecular state.
Thus, a narrow resonance state $\Lambda_c J/\psi$ with molecular configuration, whose corrected mass and decay width are 5367 and 1 MeV, is confirmed.

\begin{figure}[htb]
	\centering
	\includegraphics[width=8cm]{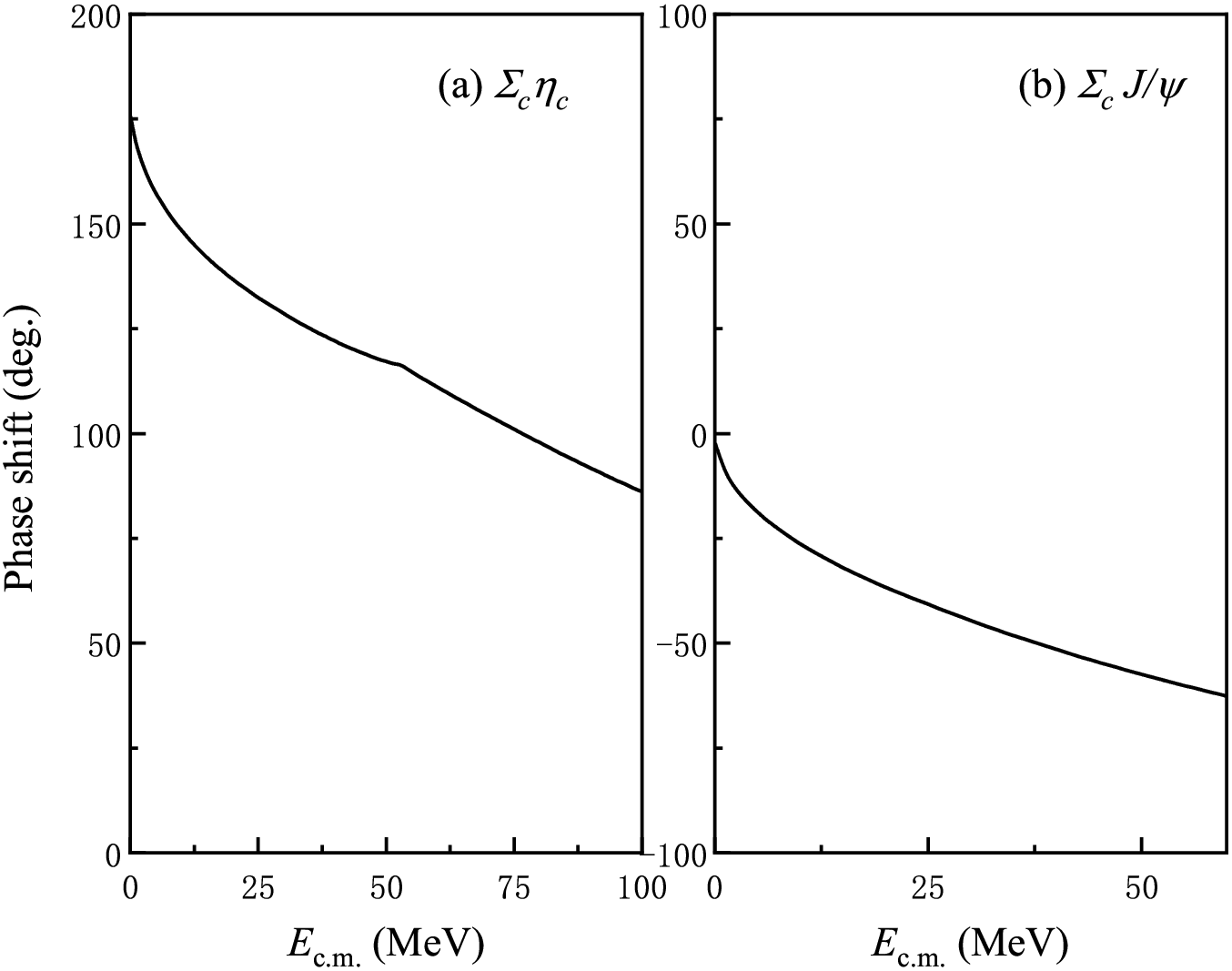}\
	\caption{\label{shift2}  The phase shifts of the open channel (a) $\Sigma_c \eta_c$ and (b) $\Sigma_c J/\psi$ with $I(J^P) = 1(\frac{1}{2}^-)$ channel coupling.}
\end{figure}

In Fig.~\ref{shift2}, we study the scattering phase shifts of the open channel $\Sigma_c \eta_c$ and $\Sigma_c J/\psi$ to examine whether the $\Sigma_c^*J/\psi$ can form a resonance state.
However, the phase shifts of open channels do not show a sharp increase around the energies of the quasibound state $\Sigma_c^*J/\psi$.
The result shows that the $\Sigma_c^*J/\psi$ becomes scattering state rather than resonance state.
We further examine the eigenvalues obtained after channel coupling and find that the lowest energy with the main component of the $\Sigma_c^*J/\psi$ is above its threshold.
This indicates that the energy of the $\Sigma_c^*J/\psi$ is elevated by the $\Sigma_c \eta_c$ and $\Sigma_c J/\psi$ in channel coupling process, thus the $\Sigma_c^*J/\psi$ becomes scattering state.
This can be understood by the fact that the binding energy of the $\Sigma_c^*J/\psi$ in the single-channel calculation is very small ($\leq$ 2 MeV).
So it can be easily pushed above the corresponding threshold after the channel coupling and becomes a scattering state.
In addition, one can notice that in both Fig.~\ref{shift1} and~\ref{shift2}, as the incident energy approaches 0 MeV, the phase shifts of the open channels $\Lambda_c \eta_c$ and $\Sigma_c \eta_c$ tend to 180$^\circ$, which conforms to the characteristics of bound states.

The obtained states in this work are summarized in Table~\ref{sum}.
Additionally, there have been some theoretical studies on the $qqc\bar{c}c$ pentaquark states~\cite{Chen:2017jjn,Wang:2019aoc,Wang:2018ihk,An:2019idk}.
In Ref.~\cite{Wang:2019aoc,Chen:2017jjn}, several possible molecular states are obtained, which are the $\Xi_{cc} \bar{D}^*$ with $I(J^P) = 0(1/2^-)$, $\Xi_{cc} \bar{D}_1$ with $I(J^P) = 0(1/2^+, 3/2^+)$, and $\Xi_{cc} \bar{D}_2^*$ with $I(J^P) = 0(3/2^+, 5/2^+)$.
However, the composition of these possible states is not the same as that studied in our present work.
In Ref.~\cite{Wang:2018ihk}, the masses of the $P_{cc\bar{c}ud}$ pentaquark states with $J^P=1/2^+$ and $J^P=1/2^-$ are calculated, which are about 5610 and 5720 MeV.
In addition, the structure of the two states is the scalar-diquark-scalar-diquark-antiquark type.
In Ref.~\cite{An:2019idk}, a stable $qqc\bar{c}c$ pentaquark state with $I(J^P) = 0(5/2^-)$ is obtained, whose mass is 5681 MeV.
Our work also leads to a similar conclusion that the $qqc\bar{c}c$ with $I(J^P) = 0(5/2^-)$ can form a pentaquark state.
Moreover, according to our calculations, it has a molecular configuration.

\begin{table}[htb]
	\caption{\label{sum} The states obtained in this work.}
	\begin{tabular}{c c c} \hline\hline
		
		~~$I(J^P)$~~  & ~~~~~~Mass (in MeV)~~~~~~ & ~~~Configuration~~~   \\ \hline
		
		$0(1/2^-)$  & 5259    & compact pentaquark   \\
		$0(1/2^-)$  & 5367    & molecular state   \\
		$0(5/2^-)$  & 5690    & molecular state \\
		$1(1/2^-)$  & 5396    & compact pentaquark   \\
		$1(3/2^-)$  & 5465    & compact pentaquark   \\ \hline\hline
	\end{tabular}
\end{table}

\section{Summary}

In this work, we investigate the hidden-charm $qqc\bar{c}c$ systems in the framework of the QDCSM.
The $S$-wave pentaquark systems with $I$ = 0 and 1, $J^P$ = $1/2^-$, $3/2^-$, and $5/2^-$ are considered.
The effective potnetial is studied to describe the interaction between baryons and mesons.
Both the single-channel and the coupled-channel dynamic bound-state calculation is carried out to search for possible states.
Meanwhile, the study of the scattering process of the open channels is carried out to confirm possible resonance states.
We also calculate the rms of cluster spacing to further determine the structure of the obtained states.

According to our numerical results, we obtain three compact $qqc\bar{c}c$ pentaquark states.
The masses and the quantum numbers of them are 5259 MeV with $I(J^P)$ = $0(1/2^-)$, 5396 MeV with $I(J^P)$ = $1(1/2^-)$, and 5465 MeV with $I(J^P)$ = $1(3/2^-)$.
In the decay channel $\Lambda_c \eta_c$, we find an $I(J^P) = 0(1/2^-)$ molecular state $\Lambda_c J/\psi$ and its mass and decay width are 5367 and 1 MeV, respectively.
Another obtained molecular state is $I(J^P)$ = $0(5/2^-)$ $\Xi_{cc}^* \bar{D}^*$, whose mass is 5690 MeV. All these states are worthy of further experimental exploration.
In addition, the present study shows that the influence of channel coupling is necessary in describing the multiquark system.
Several obtained states in this work result from the influence of channel coupling and these states tend to have a compact configuration.
Based on this, we would like to emphasize the importance of channel coupling in studying exotic hadron states.

\acknowledgments{This work is supported partly by the National Science Foundation of China under Contract Nos. and Postgraduate Research and Practice Innovation Program of Jiangsu Province under Grant no.KYCX23\underline{~}1675.}	

\setcounter{equation}{0}
\renewcommand\theequation{A\arabic{equation}}

\section*{Appendix: Resonating group method for bound-state and scattering process}

The resonating group method (RGM)~\cite{RGM1,RGM} and generating coordinates method~\cite{GCM1,GCM2} are used to carry out a dynamical calculation.
The main feature of the RGM for two-cluster systems is that it assumes that two clusters are frozen inside, and only considers the relative motion between the two clusters.
So the conventional ansatz for the two-cluster wave functions is
\begin{equation}
	\psi_{5q} = {\cal A }\left[[\phi_{B}\phi_{M}]^{[\sigma]IS}\otimes\chi(\boldsymbol{R})\right]^{J}, \label{5q}
\end{equation}
where the symbol ${\cal A }$ is the antisymmetrization operator, and ${\cal A} = 1-P_{14}-P_{24}-P_{34}$. $[\sigma]=[222]$ gives the total color symmetry and all other symbols have their usual meanings.
$\phi_{B}$ and $\phi_{M}$ are the $q^{3}$ and $\bar{q}q$ cluster wave functions, respectively.
From the variational principle, after variation with respect to the relative motion wave function $\chi(\boldsymbol{\mathbf{R}})=\sum_{L}\chi_{L}(\boldsymbol{\mathbf{R}})$, one obtains the RGM equation:
\begin{equation}
	\int H(\boldsymbol{\mathbf{R}},\boldsymbol{\mathbf{R'}})\chi(\boldsymbol{\mathbf{R'}})d\boldsymbol{\mathbf{R'}}=E\int N(\boldsymbol{\mathbf{R}},\boldsymbol{\mathbf{R'}})\chi(\boldsymbol{\mathbf{R'}})d\boldsymbol{\mathbf{R'}},  \label{RGM eq}
\end{equation}
where $H(\boldsymbol{\mathbf{R}},\boldsymbol{\mathbf{R'}})$ and $N(\boldsymbol{\mathbf{R}},\boldsymbol{\mathbf{R'}})$ are Hamiltonian and norm kernels.
By solving the RGM equation, we can get the energies $E$ and the wave functions.
In fact, it is not convenient to work with the RGM expressions.
Then, we expand the relative motion wave function $\chi(\boldsymbol{\mathbf{R}})$ by using a set of Gaussians with different centers
\begin{align}
	\chi(\boldsymbol{R}) =& \frac{1}{\sqrt{4 \pi}}\left(\frac{6}{5 \pi b^{2}}\right)^{3 / 4} \sum_{i,L,M} C_{i,L} \nonumber     \\
	&\cdot\int \exp \left[-\frac{3}{5 b^{2}}\left(\boldsymbol{R}-\boldsymbol{S}_{i}\right)^{2}\right] Y_{L,M}\left(\hat{\boldsymbol{S}}_{i}\right) d \Omega_{\boldsymbol{S}_{i}}
\end{align}
where $L$ is the orbital angular momentum between two clusters, and $\boldsymbol {S_{i}}$, $i=1,2,...,n$ are the generator coordinates, which are introduced to expand the relative motion wave function. By including the center-of-mass motion:
\begin{equation}
	\phi_{C} (\boldsymbol{R}_{C}) = (\frac{5}{\pi b^{2}})^{3/4}e^{-\frac{5\boldsymbol{R}^{2}_{C}}{2b^{2}}},
\end{equation}
the ansatz Eq.~(\ref{5q}) can be rewritten as
\begin{align}
	\psi_{5 q} =& \mathcal{A} \sum_{i,L} C_{i,L} \int \frac{d \Omega_{\boldsymbol{S}_{i}}}{\sqrt{4 \pi}} \prod_{\alpha=1}^{3} \phi_{\alpha}\left(\boldsymbol{S}_{i}\right) \prod_{\beta=4}^{5} \phi_{\beta}\left(-\boldsymbol{S}_{i}\right) \nonumber \\
	& \cdot\left[\left[\chi_{I_{1} S_{1}}\left(B\right) \chi_{I_{2} S_{2}}\left(M\right)\right]^{I S} Y_{LM}\left(\hat{\boldsymbol{S}}_{i}\right)\right]^{J} \nonumber \\
	& \cdot\left[\chi_{c}\left(B\right) \chi_{c}\left(M\right)\right]^{[\sigma]}, \label{5q2}
\end{align}
where $\chi_{I_{1}S_{1}}$ and $\chi_{I_{2}S_{2}}$ are the product of the flavor and spin wave functions, and $\chi_{c}$ is the color wave function.
These will be shown in detail later.
$\phi_{\alpha}(\boldsymbol{S}_{i})$ and $\phi_{\beta}(-\boldsymbol{S}_{i})$ are the single-particle orbital wave functions with different reference centers,
\begin{align}
	\phi_{\alpha}\left(\boldsymbol{S}_{i}\right) & = \left(\frac{1}{\pi b^{2}}\right)^{3 / 4} e^{-\frac{1}{2 b^{2}}\left(r_{\alpha}-\frac{2}{5} \boldsymbol{S}_{i}\right)^{2}}, \nonumber \\
	\phi_{\beta}\left(\boldsymbol{-S}_{i}\right) & = \left(\frac{1}{\pi b^{2}}\right)^{3 / 4} e^{-\frac{1}{2 b^{2}}\left(r_{\beta}+\frac{3}{5} \boldsymbol{S}_{i}\right)^{2}}.
\end{align}
With the reformulated ansatz Eq.~(\ref{5q2}), the RGM Eq.~(\ref{RGM eq}) becomes an algebraic eigenvalue equation:
\begin{equation}
	\sum_{j} C_{j}H_{i,j}= E \sum_{j} C_{j}N_{i,j},
\end{equation}
where $H_{i,j}$ and $N_{i,j}$ are the Hamiltonian matrix elements and overlaps, respectively.
By solving the generalized eigenproblem, we can obtain the energy and the corresponding wave functions of the pentaquark systems.

For a scattering problem, the relative wave function is expanded as
\begin{align}
\chi_{L}(\mathbf{R}) & =\sum_{i} C_{i} \frac{\tilde{u}_{L}\left(\boldsymbol{R}, \boldsymbol{S}_{i}\right)}{\boldsymbol{R}} Y_{L,M}(\hat{\boldsymbol{R}}),
\end{align}
with
\begin{align}
\tilde{u}_{L}\left(\boldsymbol{R}, \boldsymbol{S}_{i}\right) & = \left\{\begin{array}{ll}
	\alpha_{i} u_{L}\left(\boldsymbol{R}, \boldsymbol{S}_{i}\right), & \boldsymbol{R} \leq \boldsymbol{R}_{C} \\
	{\left[h_{L}^{-}(\boldsymbol{k}, \boldsymbol{R})-s_{i} h_{L}^{+}(\boldsymbol{k}, \boldsymbol{R})\right] R_{A B},} & \boldsymbol{R} \geq \boldsymbol{R}_{C}
\end{array}\right.
\end{align}
where
\begin{align}
	u_{L}\left(\boldsymbol{R}, \boldsymbol{S}_{i}\right)= & \sqrt{4 \pi}\left(\frac{6}{5 \pi b^{2}}\right)^{3 / 4} \mathbf{R} e^{-\frac{3}{5 b^{2}}\left(\boldsymbol{R}-\boldsymbol{S}_{i}\right)^{2}} \nonumber \\
	& \cdot i^{L} j_{L}\left(-i \frac{6}{5 b^{2}} S_{i}\right).
\end{align}

$h^{\pm}_L$ are the $L$th spherical Hankel functions, $k$ is the momentum of the relative motion with $k=\sqrt{2 \mu E_{i e}}$, $\mu$ is the reduced mass of two hadrons of the open channel, $E_{i e}$ is the incident energy of the relevant open channels, which can be written as $E_{i e} = E_{total} - E_{th}$, where $E_{total}$ denotes the total energy, and $E_{th}$ represents the threshold of the open channel.
$R_C$ is a cutoff radius beyond which all the strong interaction can be disregarded.
Additionally, $\alpha_i$ and $s_i$ are complex parameters that are determined by the smoothness condition at $R = R_C$ and $C_i$ satisfy $\sum_i C_i = 1$. After performing the variational procedure, a $L$th partial-wave equation for the scattering problem can be deduced as
\begin{align}
\sum_j \mathcal{L}_{i j}^L C_j &= \mathcal{M}_i^L(i=0,1, \ldots, n-1),
\end{align}
with
\begin{align}
	 \mathcal{L}_{i j}^L&=\mathcal{K}_{i j}^L-\mathcal{K}_{i 0}^L-\mathcal{K}_{0 j}^L+\mathcal{K}_{00}^L, \nonumber \\
	 \mathcal{M}_i^L&=\mathcal{K}_{00}^L-\mathcal{K}_{i 0}^L,
\end{align}
and
\begin{align}
	\mathcal{K}_{i j}^L= & \left\langle\hat{\phi}_A \hat{\phi}_B \frac{\tilde{u}_L\left(\boldsymbol{R}^{\prime}, \boldsymbol{S}_i\right)}{\boldsymbol{R}^{\prime}} Y_{L,M}\left(\boldsymbol{R}^{\prime}\right)|H-E|\right. \nonumber \\
	& \left.\cdot \mathcal{A}\left[\hat{\phi}_A \hat{\phi}_B \frac{\tilde{u}_L\left(\boldsymbol{R}, \boldsymbol{S}_j\right)}{\boldsymbol{R}} Y_{L,M}(\boldsymbol{R})\right]\right\rangle .
\end{align}
By solving Eq.~(A11), we can obtain the expansion coefficients $C_i$, then the $S$-matrix element $S_L$ and the phase shifts $\delta_L$ are given by
\begin{align}
S_L&=e^{2 i \delta_L}=\sum_{i} C_i s_i.
\end{align}

Resonances are unstable particles usually observed as bell-shaped structures in scattering cross sections of their open channels.
For a simple narrow resonance, its fundamental properties correspond to the visible cross section features: mass $M$ is at the peak position, and decay width $\Gamma$ is the half-width of the bell shape.
The cross section $\sigma_{L}$ and the scattering phase shifts $\delta_{L}$ have relations
\begin{align}
\sigma_L&=\frac{4 \pi}{k^2}(2 L+1) \sin ^2 \delta_L.
\end{align}
Therefore, resonances can also usually be observed in the scattering phase shift, where the phase shift of the scattering channels rises through $\pi/2$ at a resonance mass.
We can obtain a resonance mass at the position of the phase shift of $\pi/2$.
The decay width is the mass difference between the phase shift of $3\pi/4$ and $\pi/4$.

\setcounter{equation}{0}
\renewcommand\theequation{B\arabic{equation}}


\begin{thebibliography}{99}
	
\bibitem{LHCbPc2015} Aaij R., {\it et al.} (LHCb Collaboration),  Phys. Rev. Lett. {\bf 115} 072001 (2015).
\bibitem{LHCbPc2019} Aaij R., {\it et al.}, (LHCb Collaboration),  Phys. Rev. Lett. {\bf 122} 222001 (2019).
\bibitem{Wu:2010jy} J.~J.~Wu, R.~Molina, E.~Oset and B.~S.~Zou, Phys. Rev. Lett. \textbf{105}, 232001 (2010).
\bibitem{Wang:2011rga} W.~L.~Wang, F.~Huang, Z.~Y.~Zhang and B.~S.~Zou, Phys. Rev. C \textbf{84}, 015203 (2011).
\bibitem{Yang:2011wz} Z.~C.~Yang, Z.~F.~Sun, J.~He, X.~Liu and S.~L.~Zhu, Chin. Phys. C \textbf{36}, 6 (2012).
\bibitem{Xiao:2013yca} C.~W.~Xiao, J.~Nieves and E.~Oset, Phys. Rev. D \textbf{88}, 056012 (2013).
\bibitem{Chen:2015moa} H.~X.~Chen, W.~Chen, X.~Liu, T.~G.~Steele and S.~L.~Zhu, Phys. Rev. Lett. \textbf{115}, 172001 (2015).
\bibitem{Wang:2015epa} Z.~G.~Wang, Eur. Phys. J. C \textbf{76}, 70 (2016).
\bibitem{Wang:2015ava} Z.~G.~Wang and T.~Huang, Eur. Phys. J. C \textbf{76}, 43 (2016).
\bibitem{Azizi:2016dhy} K.~Azizi, Y.~Sarac and H.~Sundu, Phys. Rev. D \textbf{95}, 094016 (2017).
\bibitem{Azizi:2018bdv} K.~Azizi, Y.~Sarac and H.~Sundu, Phys. Lett. B \textbf{782}, 694 (2018).
\bibitem{Chen:2019bip} H.~X.~Chen, W.~Chen and S.~L.~Zhu, Phys. Rev. D \textbf{100}, 051501 (2019).
\bibitem{Zhang:2019xtu} J.~R.~Zhang, Eur. Phys. J. C \textbf{79}, 1001 (2019).
\bibitem{Xu:2019zme} Y.~J.~Xu, C.~Y.~Cui, Y.~L.~Liu and M.~Q.~Huang, Phys. Rev. D \textbf{102},034028 (2020).
\bibitem{Wang:2019got} Z.~G.~Wang, Int. J. Mod. Phys. A \textbf{35}, 2050003 (2020).
\bibitem{Azizi:2020ogm} K.~Azizi, Y.~Sarac and H.~Sundu, Chin. Phys. C \textbf{45}, 053103 (2021).
\bibitem{Xu:2020flp} Y.~J.~Xu, Y.~L.~Liu and M.~Q.~Huang, Eur. Phys. J. C \textbf{81}, 421 (2021).
\bibitem{Ozdem:2018qeh} U.~\"Ozdem and K.~Azizi, Eur. Phys. J. C \textbf{78}, 379 (2018).
\bibitem{Lu:2015fva} Q.~F.~L\"u, X.~Y.~Wang, J.~J.~Xie, X.~R.~Chen and Y.~B.~Dong, Phys. Rev. D \textbf{93}, 034009 (2016).
\bibitem{Lu:2016nnt} Q.~F.~L\"u and Y.~B.~Dong, Phys. Rev. D \textbf{93}, 074020 (2016).
\bibitem{Xiao:2019mvs} C.~J.~Xiao, Y.~Huang, Y.~B.~Dong, L.~S.~Geng and D.~Y.~Chen, Phys. Rev. D \textbf{100}, 014022 (2019).
\bibitem{Wu:2019rog} Q.~Wu and D.~Y.~Chen, Phys. Rev. D \textbf{100}, 114002 (2019).
\bibitem{Wang:2019dsi} X.~Y.~Wang, J.~He, X.~R.~Chen, Q.~Wang and X.~Zhu, Phys. Lett. B \textbf{797}, 134862 (2019).
\bibitem{Zhang:2020erj} B.~T.~Zhang, J.~S.~Wang and Y.~L.~Ma, [arXiv:2002.10954 [hep-ph]].
\bibitem{Xu:2020gjl} H.~Xu, Q.~Li, C.~H.~Chang and G.~L.~Wang, Phys. Rev. D \textbf{101}, 054037 (2020).
\bibitem{Li:2023zag} Q.~Li, C.~H.~Chang, T.~Wang and G.~L.~Wang, JHEP \textbf{06}, 189 (2023).
\bibitem{Roca:2015dva} L.~Roca, J.~Nieves and E.~Oset, Phys. Rev. D \textbf{92}, 094003 (2015).
\bibitem{Shimizu:2016rrd} Y.~Shimizu, D.~Suenaga and M.~Harada, Phys. Rev. D \textbf{93}, 114003 (2016).
\bibitem{Meng:2019ilv} L.~Meng, B.~Wang, G.~J.~Wang and S.~L.~Zhu, Phys. Rev. D \textbf{100}, 014031 (2019).
\bibitem{He:2015cea} J.~He, Phys. Lett. B \textbf{753}, 547 (2016).
\bibitem{He:2019ify} J.~He, Eur. Phys. J. C \textbf{79}, 393 (2019).
\bibitem{Oset:2016nvf} E.~Oset, H.~X.~Chen, A.~Feijoo, L.~S.~Geng, W.~H.~Liang, D.~M.~Li, J.~X.~Lu, V.~K.~Magas, J.~Nieves and A.~Ramos, \textit{et al.} Nucl. Phys. A \textbf{954}, 371 (2016).
\bibitem{Wang:2019ato} B.~Wang, L.~Meng and S.~L.~Zhu, JHEP \textbf{11}, 108 (2019).
\bibitem{Peng:2020xrf} F.~Z.~Peng, M.~Z.~Liu, M.~S\'anchez S\'anchez and M.~Pavon Valderrama, Phys. Rev. D \textbf{102}, 114020 (2020).
\bibitem{Wang:2015qlf} G.~J.~Wang, L.~Ma, X.~Liu and S.~L.~Zhu, Phys. Rev. D \textbf{93}, 034031 (2016).
\bibitem{Ali:2016dkf} A.~Ali, I.~Ahmed, M.~J.~Aslam and A.~Rehman, Phys. Rev. D \textbf{94}, 054001 (2016).
\bibitem{Shimizu:2019ptd} Y.~Shimizu, Y.~Yamaguchi and M.~Harada, [arXiv:1904.00587 [hep-ph]].
\bibitem{Du:2021fmf} M.~L.~Du, V.~Baru, F.~K.~Guo, C.~Hanhart, U.~G.~Mei\ss{}ner, J.~A.~Oller and Q.~Wang, JHEP \textbf{08}, 157 (2021).
\bibitem{Guo:2015umn} F.~K.~Guo, U.~G.~Mei\ss{}ner, W.~Wang and Z.~Yang, Phys. Rev. D \textbf{92}, 071502 (2015).
\bibitem{Liu:2015fea} X.~H.~Liu, Q.~Wang and Q.~Zhao, Phys. Lett. B \textbf{757}, 231 (2016).
\bibitem{Guo:2016bkl} F.~K.~Guo, U.~G.~Mei\ss{}ner, J.~Nieves and Z.~Yang, Eur. Phys. J. A \textbf{52}, 318 (2016).
\bibitem{Santopinto:2016pkp} E.~Santopinto and A.~Giachino, Phys. Rev. D \textbf{96}, 014014 (2017).
\bibitem{Ortega:2016syt} P.~G.~Ortega, D.~R.~Entem and F.~Fern\'andez, Phys. Lett. B \textbf{764}, 207 (2017).
\bibitem{Park:2017jbn} W.~Park, A.~Park, S.~Cho and S.~H.~Lee, Phys. Rev. D \textbf{95}, 054027 (2017).
\bibitem{Zhu:2019iwm} R.~Zhu, X.~Liu, H.~Huang and C.~F.~Qiao, Phys. Lett. B \textbf{797}, 134869 (2019).
\bibitem{Phumphan:2021tta} K.~Phumphan, K.~Xu, W.~Ruangyoo, C.~C.~Chen, A.~Limphirat and Y.~Yan, [arXiv:2105.03150 [hep-ph]].
\bibitem{Yang:2015bmv} G.~Yang and J.~Ping, Phys. Rev. D \textbf{95}, 014010 (2017).
\bibitem{Dong:2020nwk} Y.~Dong, P.~Shen, F.~Huang and Z.~Zhang, Eur. Phys. J. C \textbf{80}, 341 (2020).
\bibitem{HuangPc1} H. X. Huang, C. R. Deng, J. L. Ping and F. Wang, Eur. Phys. J. C {\bf 76}, 624 (2016).
\bibitem{HuangPc2} H. X. Huang and J. L. Ping, Phys. Rev. D {\bf 99}, 014010 (2019).
\bibitem{Zhu:2023hyh} X.~Zhu, Y.~Wu, H.~Huang, J.~Ping and Y.~Yang, Universe \textbf{9}, 265 (2023).


\bibitem{Chen:2015loa} R.~Chen, X.~Liu, X.~Q.~Li and S.~L.~Zhu, Phys. Rev. Lett. \textbf{115}, 132002 (2015).
\bibitem{Liu:2019zvb} M.~Z.~Liu, T.~W.~Wu, M.~S\'anchez S\'anchez, M.~P.~Valderrama, L.~S.~Geng and J.~J.~Xie, Phys. Rev. D \textbf{103}, 054004 (2021).
\bibitem{Chen:2019asm} R.~Chen, Z.~F.~Sun, X.~Liu and S.~L.~Zhu, Phys. Rev. D \textbf{100}, 011502 (2019).
\bibitem{Yalikun:2021bfm} N.~Yalikun, Y.~H.~Lin, F.~K.~Guo, Y.~Kamiya and B.~S.~Zou, Phys. Rev. D \textbf{104}, 094039 (2021).


\bibitem{Deng:2016rus} C.~Deng, J.~Ping, H.~Huang and F.~Wang, Phys. Rev. D \textbf{95}, 014031 (2017).

\bibitem{Scoccola:2015nia} N.~N.~Scoccola, D.~O.~Riska and M.~Rho, Phys. Rev. D \textbf{92}, 051501 (2015).
\bibitem{Wu:2017weo} J.~Wu, Y.~R.~Liu, K.~Chen, X.~Liu and S.~L.~Zhu, Phys. Rev. D \textbf{95}, 034002 (2017).
\bibitem{Weng:2019ynv} X.~Z.~Weng, X.~L.~Chen, W.~Z.~Deng and S.~L.~Zhu, Phys. Rev. D \textbf{100}, 016014 (2019).
\bibitem{Zhu:2015bba} R.~Zhu and C.~F.~Qiao, Phys. Lett. B \textbf{756}, 259 (2016).
\bibitem{Wang:2016dzu} G.~J.~Wang, R.~Chen, L.~Ma, X.~Liu and S.~L.~Zhu, Phys. Rev. D \textbf{94}, 094018 (2016).
\bibitem{Hiyama:2018ukv} E.~Hiyama, A.~Hosaka, M.~Oka and J.~M.~Richard, Phys. Rev. C \textbf{98}, 045208 (2018).
\bibitem{Burns:2019iih} T.~J.~Burns and E.~S.~Swanson, Phys. Rev. D \textbf{100}, 114033 (2019).
\bibitem{Yang:2022ezl} Z.~Y.~Yang, F.~Z.~Peng, M.~J.~Yan, M.~S\'anchez S\'anchez and M.~Pavon Valderrama, [arXiv:2211.08211 [hep-ph]].
\bibitem{Meissner:2015mza} U.~G.~Mei\ss{}ner and J.~A.~Oller, Phys. Lett. B \textbf{751}, 59 (2015).
\bibitem{Mutuk:2019snd} H.~Mutuk, Chin. Phys. C \textbf{43}, 093103 (2019).
\bibitem{Yamaguchi:2019seo} Y.~Yamaguchi, H.~Garc\'\i{}a-Tecocoatzi, A.~Giachino, A.~Hosaka, E.~Santopinto, S.~Takeuchi and M.~Takizawa, Phys. Rev. D \textbf{101}, 091502 (2020).
\bibitem{Kubarovsky:2015aaa} V.~Kubarovsky and M.~B.~Voloshin, Phys. Rev. D \textbf{92}, 031502 (2015)/
\bibitem{Wang:2015jsa} Q.~Wang, X.~H.~Liu and Q.~Zhao, Phys. Rev. D \textbf{92}, 034022 (2015).
\bibitem{HillerBlin:2016odx} A.~N.~Hiller Blin, C.~Fern\'andez-Ram\'\i{}rez, A.~Jackura, V.~Mathieu, V.~I.~Mokeev, A.~Pilloni and A.~P.~Szczepaniak, Phys. Rev. D \textbf{94}, 034002 (2016).
\bibitem{Paryev:2018fyv} E.~Y.~Paryev and Y.~T.~Kiselev, Nucl. Phys. A \textbf{978}, 201 (2018).
\bibitem{Guo:2019fdo} F.~K.~Guo, H.~J.~Jing, U.~G.~Mei\ss{}ner and S.~Sakai, Phys. Rev. D \textbf{99},091501 (2019).
\bibitem{Shen:2016tzq} C.~W.~Shen, F.~K.~Guo, J.~J.~Xie and B.~S.~Zou, Nucl. Phys. A \textbf{954}, 393 (2016).
\bibitem{Burns:2015dwa} T.~J.~Burns, Eur. Phys. J. A \textbf{51}, 152 (2015).
\bibitem{ChenHX0} Chen H. X., Chen W., Liu X. and Zhu S. L., {\it Phys. Rep.} {\bf 639} 1 (2016).
\bibitem{LiuYR0} Liu Y. R., Chen H. X., Chen W., Liu X. and Zhu S. L., {\it Progress in Particle and Nuclear Physics.}, {\bf 107} 237 (2019).
\bibitem{YangG0} Yang G., Ping J. L. and Segovia J., {\it Symmetry.}, {\bf 12} 1869 (2020).
\bibitem{LHCb:2020jpq} R.~Aaij \textit{et al.} [LHCb], Sci. Bull. \textbf{66}, 1278 (2021).
\bibitem{LHCb:2022ogu} R.~Aaij \textit{et al.} [LHCb], Phys. Rev. Lett. \textbf{131}, 031901 (2023).
\bibitem{Chen:2020uif} H.~X.~Chen, W.~Chen, X.~Liu and X.~H.~Liu, Eur. Phys. J. C \textbf{81}, 409 (2021).
\bibitem{Wang:2020eep} Z.~G.~Wang, Int. J. Mod. Phys. A \textbf{36}, 2150071 (2021).
\bibitem{Azizi:2021utt} K.~Azizi, Y.~Sarac and H.~Sundu, Phys. Rev. D \textbf{103}, 094033 (2021).
\bibitem{Wang:2021itn} Z.~G.~Wang and Q.~Xin, Chin. Phys. C \textbf{45}, 123105 (2021).
\bibitem{Wang:2022neq} X.~W.~Wang and Z.~G.~Wang, Chin. Phys. C \textbf{47}, 013109 (2023).
\bibitem{Azizi:2023iym} K.~Azizi, Y.~Sarac and H.~Sundu, Phys. Rev. D \textbf{108}, 074010 (2023).
\bibitem{Ozdem:2021ugy} U.~\"Ozdem, Eur. Phys. J. C \textbf{81}, 277 (2021).
\bibitem{Ozdem:2022kei} U.~\"Ozdem, Phys. Lett. B \textbf{836}, 137635 (2023).
\bibitem{Ozdem:2023htj} U.~\"Ozdem, [arXiv:2303.10649 [hep-ph]].
\bibitem{Peng:2020hql} F.~Z.~Peng, M.~J.~Yan, M.~S\'anchez S\'anchez and M.~P.~Valderrama, Eur. Phys. J. C \textbf{81}, 666 (2021).
\bibitem{Liu:2020hcv} M.~Z.~Liu, Y.~W.~Pan and L.~S.~Geng, Phys. Rev. D \textbf{103}, 034003 (2021).
\bibitem{Yan:2021nio} M.~J.~Yan, F.~Z.~Peng, M.~S\'anchez S\'anchez and M.~Pavon Valderrama, Eur. Phys. J. C \textbf{82}, 574 (2022).
\bibitem{Yan:2022wuz} M.~J.~Yan, F.~Z.~Peng, M.~S\'anchez S\'anchez and M.~Pavon Valderrama, Phys. Rev. D \textbf{107}, 074025 (2023).
\bibitem{Cheng:2021gca} C.~Cheng, F.~Yang and Y.~Huang, Phys. Rev. D \textbf{104}, 116007 (2021).
\bibitem{Yang:2021pio} F.~Yang, Y.~Huang and H.~Q.~Zhu, Sci. China Phys. Mech. Astron. \textbf{64}, 121011 (2021).
\bibitem{Wu:2021caw} Q.~Wu, D.~Y.~Chen and R.~Ji, Chin. Phys. Lett. \textbf{38}, 071301 (2021).
\bibitem{Clymton:2021thh} S.~Clymton, H.~J.~Kim and H.~C.~Kim, Phys. Rev. D \textbf{104}, 014023 (2021).
\bibitem{Zhu:2021lhd} J.~T.~Zhu, L.~Q.~Song and J.~He, Phys. Rev. D \textbf{103}, 074007 (2021).
\bibitem{Zhu:2022wpi} J.~T.~Zhu, S.~Y.~Kong and J.~He, Phys. Rev. D \textbf{107}, 034029 (2023).
\bibitem{Burns:2022uha} T.~J.~Burns and E.~S.~Swanson, Phys. Lett. B \textbf{838}, 137715 (2023).
\bibitem{Feijoo:2022rxf} A.~Feijoo, W.~F.~Wang, C.~W.~Xiao, J.~J.~Wu, E.~Oset, J.~Nieves and B.~S.~Zou, Phys. Lett. B \textbf{839}, 137760 (2023).
\bibitem{Xiao:2021rgp} C.~W.~Xiao, J.~J.~Wu and B.~S.~Zou, Phys. Rev. D \textbf{103}, 054016 (2021).
\bibitem{Wang:2022tib} F.~L.~Wang, H.~Y.~Zhou, Z.~W.~Liu and X.~Liu, Phys. Rev. D \textbf{106}, 054020 (2022).
\bibitem{Ortega:2022uyu} P.~G.~Ortega, D.~R.~Entem and F.~Fernandez, Phys. Lett. B \textbf{838}, 137747 (2023).
\bibitem{Gao:2021hmv} F.~Gao and H.~S.~Li, Chin. Phys. C \textbf{46}, 123111 (2022).
\bibitem{Hu:2021nvs} X.~Hu and J.~Ping, Eur. Phys. J. C \textbf{82}, 118 (2022).
\bibitem{Chen:2020kco} R.~Chen, Phys. Rev. D \textbf{103}, 054007 (2021).
\bibitem{Chen:2021tip} R.~Chen, Eur. Phys. J. C \textbf{81}, 122 (2021).
\bibitem{Wang:2022mxy} F.~L.~Wang and X.~Liu, Phys. Lett. B \textbf{835}, 137583 (2022).
\bibitem{Chen:2022onm} R.~Chen and X.~Liu, Phys. Rev. D \textbf{105}, 014029 (2022).
\bibitem{Li:2023aui} S.~Y.~Li, Y.~R.~Liu, Z.~L.~Man, Z.~G.~Si and J.~Wu, Phys. Rev. D \textbf{108}, 056015 (2023).
\bibitem{Deng:2022vkv} C.~R.~Deng, Phys. Rev. D \textbf{105}, 116021 (2022).
\bibitem{Meng:2022wgl} L.~Meng, B.~Wang and S.~L.~Zhu, Phys. Rev. D \textbf{107}, 014005 (2023).
\bibitem{Giachino:2022pws} A.~Giachino, A.~Hosaka, E.~Santopinto, S.~Takeuchi, M.~Takizawa and Y.~Yamaguchi, Phys. Rev. D \textbf{108}, 074012 (2023).
\bibitem{Paryev:2022zdx} E.~Y.~Paryev, Nucl. Phys. A \textbf{1023}, 122452 (2022).
\bibitem{Chen:2022wkh} K.~Chen, Z.~Y.~Lin and S.~L.~Zhu, Phys. Rev. D \textbf{106}, 116017 (2022).
\bibitem{Paryev:2023icm} E.~Y.~Paryev, Nucl. Phys. A \textbf{1037}, 122687 (2023).
\bibitem{Nakamura:2022gtu} S.~X.~Nakamura and J.~J.~Wu, Phys. Rev. D \textbf{108}, L011501 (2023).
\bibitem{Liu:2020ajv} W.~Y.~Liu, W.~Hao, G.~Y.~Wang, Y.~Y.~Wang, E.~Wang and D.~M.~Li, Phys. Rev. D \textbf{103}, 034019 (2021).
\bibitem{Gershon:2022xnn} T.~Gershon [LHCb], [arXiv:2206.15233 [hep-ex]].
\bibitem{Hofmann:2005sw} J.~Hofmann and M.~F.~M.~Lutz, Nucl. Phys. A \textbf{763}, 90 (2005).
\bibitem{Chen:2017jjn} R.~Chen, A.~Hosaka and X.~Liu, Phys. Rev. D \textbf{96}, 114030 (2017).
\bibitem{Wang:2018ihk} Z.~G.~Wang, Eur. Phys. J. C \textbf{78}, 300 (2018).
\bibitem{Wang:2019aoc} F.~L.~Wang, R.~Chen, Z.~W.~Liu and X.~Liu, Phys. Rev. D \textbf{99}, 054021 (2019).
\bibitem{An:2019idk} H.~T.~An, Q.~S.~Zhou, Z.~W.~Liu, Y.~R.~Liu and X.~Liu, Phys. Rev. D \textbf{100}, 056004 (2019).
\bibitem{Wu:1996fm} G.~H.~Wu, L.~J.~Teng, J.~L.~Ping, F.~Wang and J.~T.~Goldman, Phys. Rev. C {\bf 53}, 1161 (1996).
\bibitem{Ping:1998si} J.~L.~Ping, F.~Wang and J.~T.~Goldman, Nucl. Phys. A {\bf 657}, 95 (1999).
\bibitem{Wu:1998wu} G.~h.~Wu, J.~L.~Ping, L.~j.~Teng, F.~Wang and J.~T.~Goldman, Nucl. Phys. A {\bf 673}, 279 (2000).
\bibitem{Pang:2001xx} H.~R.~Pang, J.~L.~Ping, F.~Wang and J.~T.~Goldman, Phys. Rev. C {\bf 65}, 014003 (2002).
\bibitem{Ping:2000dx} J.~L.~Ping, F.~Wang and J.~T.~Goldman, Phys. Rev. C {\bf 65}, 044003 (2002).
\bibitem{Huang:2023jec} H.~Huang, C.~Deng, X.~Liu, Y.~Tan and J.~Ping, Symmetry \textbf{15}, 1298 (2023).
\bibitem{ChenLZ} L. Z. Chen, H. R. Pang, H. X. Huang, J. L. Ping and F. Wang, Phys. Rev. C {\bf 76}, 014001 (2007).	
\bibitem{Huang:2011kf} H.~Huang, P.~Xu, J.~Ping and F.~Wang, Phys. Rev. C {\bf 84}, 064001 (2011).
\bibitem{Yan:2022nxp} Y.~Yan, X.~Hu, Y.~Wu, H.~Huang, J.~Ping and Y.~Yang, Eur. Phys. J. C \textbf{83}, 524 (2023).
\bibitem{DeRujula:1975qlm} A.~De Rujula, H.~Georgi and S.~L.~Glashow, Phys. Rev. D {\bf 12}, 147 (1975).
\bibitem{Isgur:1978xj} N.~Isgur and G.~Karl, Phys. Rev. D {\bf 18}, 4187 (1978).
\bibitem{Isgur:1978wd} N.~Isgur and G.~Karl, Phys. Rev. D {\bf 19}, 2653 (1979).
\bibitem{Isgur:1979be} N.~Isgur and G.~Karl, Phys. Rev. D {\bf 20}, 1191 (1979).
\bibitem{Ping:2000cb} J.~l.~Ping, F.~Wang and J.~T.~Goldman, Nucl. Phys. A {\bf 688}, 871 (2001).
\bibitem{Ping:2008tp} J.~L.~Ping, H.~X.~Huang, H.~R.~Pang, F.~Wang and C.~W.~Wong, Phys. Rev. C {\bf 79}, 024001 (2009).
\bibitem{ChenM} M. Chen, H. X. Huang, J. L. Ping and F. Wang, Phys. Rev. C {\bf 83}, 015202 (2011).
\bibitem{Workman:2022ynf} R.~L.~Workman \textit{et al.} [Particle Data Group], PTEP \textbf{2022}, 083C01 (2022).
\bibitem{Hu:2022pae} X.~Hu and J.~Ping, Phys. Rev. D \textbf{106}, 054028 (2022).
\bibitem{Yan:2022nxp} Y.~Yan, X.~Hu, Y.~Wu, H.~Huang, J.~Ping and Y.~Yang, Eur. Phys. J. C \textbf{83}, 524 (2023).
\bibitem{Ortiz-Pacheco:2023kjn} E.~Ortiz-Pacheco and R.~Bijker, Phys. Rev. D \textbf{108}, 054014 (2023).
\bibitem{Xu} M. M. Xu, M. Yu and L. S. Liu, Phys. Rev. Lett. {\bf 100}, 092301 (2008).
\bibitem{Xia:2021tof} Z.~Xia, S.~Fan, X.~Zhu, H.~Huang and J.~Ping, Phys. Rev. C \textbf{105}, 025201 (2022).
\bibitem{Chen:2011zzb} M.~Chen, H.~Huang, J.~Ping and F.~Wang, Phys. Rev. C \textbf{83}, 015202 (2011).
\bibitem{RGM1} J. A. Wheeler, Phys. Rev. {\bf 52}, 1083 (1937).
\bibitem{RGM} M. Kamimura, Prog. Theor. Phys. Suppl. {\bf 62}, 236 (1977).
\bibitem{GCM1} D. L. Hill and J. A. Wheeler, Phys. Rev. {\bf 89}, 1102 (1953).
\bibitem{GCM2} J. J. Griffin and J. A. Wheeler, Phys. Rev. {\bf 108}, 311 (1957).


\end{thebibliography}
\end{document}